\documentclass[sigconf,nonacm]{acmart}

\renewcommand\footnotetextcopyrightpermission[1]{} 
\usepackage[english]{babel}
\usepackage{subcaption}

\renewcommand\footnotetextcopyrightpermission[1]{} 
\setcopyright{none}

\begin{document}
\title{A Design for an Early Quantum Network}


	\author{Yuan Li, Chen Zhang,Hao Zhang,  Tao Huang, Yunjie Liu}
	
\thanks{(Corresponding author: Yuan Li)}
\thanks{Yuan Li,  Chen Zhang, and Hao Zhang are with the Future Network Research Center, Purple Mountain Laboratories, Nanjing 211111, China (e-mail: liyuan@pmlabs.com.cn).}
\thanks{Tao Huang and Yunjie Liu are with the State Key Laboratory of Networking and Switching Technology, Beijing University of Posts and Telecommunications, Beijing 100876, China, and also with the Future Network Research Center, Purple Mountain Laboratories, Nanjing 211111, China.}


\renewcommand{\shortauthors}{X.et al.}

\begin{abstract}
With the rapid advancement of quantum information technology, quantum networks have become essential for supporting diverse applications, which often have stringent demands for key metrics such as fidelity and request completion time.
In this work, we propose a design for early-stage quantum networks that is compatible with the three existing quantum repeater technologies. The design aims to maximize the ability of the network to accommodate the diverse needs of quantum applications, even under conditions of limited quantum resources and suboptimal network performance. We have also described the required identifiers in the quantum network and the specific process for implementing quantum requests.
To assess the feasibility of our design, we conduct simulations based on discrete-event modeling of quantum networks. The simulations consider various types of noise and imperfect parameters that might exist in early-stage networks. We analyze the impact of these parameters on the fidelity of the generated entangled states and the request completion time. Furthermore, we investigated additional decisions that the central controller can make beyond path selection, such as the choice of cutoff time and the allocation of network resources to requests.
\end{abstract}

\maketitle

\section{Introduction}
Currently, the fields of quantum computing\cite{steane_quantum_1998,bennett_quantum_2000,ladd_quantum_2010,nielsen_quantum_2010,preskill_quantum_2018,gyongyosi_survey_2019}, quantum communication\cite{gisin_quantum_2007}, and quantum sensing\cite{degen_quantum_2017,pirandola_advances_2018} are advancing and demonstrating their significant potential. When multiple quantum devices are interconnected through a network, their capabilities will be further enhanced.

In the realm of quantum computing, certain algorithms, specifically Shor's algorithm\cite{shor_algorithms_1994} for factoring large numbers, require a large number of quantum bits (qubits). These tasks cannot be accomplished by a quantum computer with only a few qubits. However, by connecting multiple quantum computers, these tasks become possible to be solved using distributed quantum computing\cite{buhrman_distributed_2003,beals_efficient_2013,van_meter_path_2016,cuomo_towards_2020}. For quantum communication, a network allows quantum communication between any users within the network\cite{dynes_cambridge_2019,chen_integrated_2021,chehimi_physics-informed_2022}. In the realm of quantum sensing, the connection of multiple quantum sensing devices for measurement and computation can substantially enhance measurement precision\cite{proctor_multiparameter_2018,eldredge_optimal_2018,guo_distributed_2020,zhang_distributed_2021,malia_distributed_2022}.

Nevertheless, traditional networks are incapable of supporting these functions, as the network connecting quantum devices must possess the capability to transmit quantum states, which is referred to as a quantum network\cite{kimble_quantum_2008,Dur_towards_2017,wehner_quantum_2018,kozlowski_towards_2019,azuma_tools_2021,singh_quantum_2021,ruf_quantum_2021,wei_towards_2022,illiano_quantum_2022,fang_quantum_2022,li_entanglement_2023}. Qubits exceptionally delicate and prone to losses, thus requiring relay technologies for long-distance transmission. However, the distinctive inherent non- cloning property of qubits, makes traditional relay schemes impractical. Fortunately, three generations of quantum repeater technologies\cite{munro_inside_2015,muralidharan_optimal_2016} have been proposed, offering the potential to achieve long-distance qubit transmission. Each of these three quantum repeater technologies has its own advantages and drawbacks, and the future of quantum networking is filled with uncertainty regarding whether all three technologies will coexist or if one will dominate the market.

Although quantum networks have not been fully realized, it is essential to undertake comprehensive design efforts to guide the planning of future quantum networks. Currently, a multitude of articles\cite{van_meter_recursive_2011,van_meter_path_2013,dahlberg_link_2019,hahn_quantum_2019,pant_routing_2019,shi_concurrent_2020,kozlowski_designing_2020,li_effective_2021,li_building_2021,pompili_experimental_2022,li_fidelity-guaranteed_2022,liu_quantum_2022} are exploring this field. Most of these articles focus on specific issues within quantum networks, such as routing problems. In these articles, simplifying assumptions are usually employed, such as relatively straightforward network topologies, the utilization of a fixed generation of quantum repeater technology, the absence of differentiation between users and quantum routers, and the assumption of complete knowledge regarding the quantum resource status of all network nodes. Some studies have focused on the architecture of quantum networks. For instance, \cite{van_meter_recursive_2011} proposes a quantum network scheme consisting of Q-LANs and a core network, ,\cite{li_building_2021} introduces the concept of a quantum recursive network architecture, and \cite{he_building_2024} designs a hierarchical quantum Internet architecture. However, these architectures are based on entanglement generation and do not incorporate third-generation quantum repeater schemes.

Furthermore, we anticipate that future quantum networks will have the capacity to support a range of quantum applications, including quantum computing, quantum communication, quantum sensing, etc. Different quantum requests will exhibit distinct network performance requirements, such as the desired entanglement expression, minimum acceptable end-to-end fidelity, request completion time, and more. To ensure the success of the quantum request, the network must fully comprehend these quantum request demands and meet them. Failure to meet these requirements will result in the failure of the quantum request. Therefore, the construction of a quantum network cannot follow the classical network principle of ``best-effort''.

This paper presents several key contributions and innovations:

\textbf{(1) Adaptive Quantum Network Design:} We design a quantum network tailored for early-stage deployments, addressing constraints such as scarce quantum resources and suboptimal performance parameters. Our approach ensures compatibility with the three existing quantum repeater technologies. By leveraging a central controller, we dynamically allocate network resources and configure routing strategies based on diverse quantum application demands, enabling a general-purpose quantum network rather than a dedicated network for a specific application.

\textbf{(2)	Identification Framework for Quantum Networks:} We define essential identifiers required for operation within a quantum network.

\textbf{(3)	Quantum Request Execution Workflow:} We propose a workflow for processing quantum requests, ensuring seamless execution within the network.

\textbf{(4)	Simulation-Based Analysis:} Using a quantum network simulator, we conduct analyses of key parameters affecting entanglement fidelity and request completion time. Notably, beyond accessing standard network parameters (e.g., node capabilities), we also investigate central controller decision-making factors, such as cutoff time and resource allocated for the request, and their impact on network performance.


\section{Characteristics of Quantum Networks}
Classical networks have undergone substantial maturation in their development. If the quantum network exhibits minimal differences from classical networks, it is possible to directly apply various designs from classical networks or only require minor adjustments. However, quantum networks manifest many characteristics that significantly distinguish them from classical networks, thereby demanding entirely novel solutions.  The following are some key features that distinguish quantum networks from classical networks:

\textbf{(1) Qubits possess properties such as superposition, non-locality, no-cloning, and decoherence.}\cite{wootters_single_1982,brunner_bell_2014,schlosshauer_quantum_2019}

Superposition provides quantum computing with inherent parallel computing capabilities.

Quantum entanglement\cite{horodecki_quantum_2009,bengtsson_geometry_2017}, demonstrating the non-local properties of qubits, has brought various quantum applications. Certain quantum operations associated with entanglement, including entanglement swapping\cite{zzukowski_event-ready-detectors_1993,pan_experimental_1998}, entanglement purification\cite{yan_advances_2023}, and quantum teleportation\cite{bennett_teleporting_1993,pirandola_advances_2015}, introduce new solutions to quantum networks.

No-cloning means that the common technique of copying and retransmitting in classical network is inapplicable in quantum network. Although some have suggested employing quantum secret sharing schemes for quantum retransmission\cite{yu_protocols_2021}, this approach prolongs the time required for successful transmission, and any retransmission would likewise result in extended transmission time. Given that quantum applications usually demand low latency, avoiding retransmissions as much as possible is necessary.

Decoherence indicates that qubits are very fragile and susceptible to environmental influences. A qubit could undergo decoherence in a very short time, rendering it unusable by losing its quantum properties. This emphasizes the importance of reducing latency and packet loss in quantum networks. The network must implement entanglement purification or quantum error correction\cite{devitt_quantum_2013} to preserve the quality of qubits. These operations cannot be solely carried out at the user end, as qubits would become unusable without proper protection in the network. The transfer of quantum states will also degrade their quality. Therefore, it's essential to minimize unnecessary transfer operations.

These unique quantum characteristics challenge traditional networking approaches and necessitate the development of novel solutions tailored to quantum networks.

\textbf{(2) Several quantum repeater (QR) technologies have been proposed for transmitting quantum data.} For example, the first- and second-generation QR technologies enable quantum teleportation by establishing remote entanglement. Specifically, the first-generation QR technology relies on entanglement purification to enhance the quality of qubits, while the second-generation QR technology employs quantum error correction for the same purpose. Moreover, there is the third-generation QR technology that uses quantum error correction for hop-by-hop forwarding of quantum data.

\textbf{(3) Small Scale of Quantum Networks.} Owing to the considerable challenges in the development of quantum devices, it is foreseeable that the scale of quantum networks will likely remain limited in the near future, potentially consisting of only a few hundred quantum devices. This is markedly different from classical networks, which currently encompass hundreds of millions of devices. 


Furthermore, each device in the quantum network has limited quantum memory capacity. In the absence of appropriate design measures, such networks can easily become prone to congestion issues. Congestion occurs in a quantum network when the available quantum resources become insufficient for entanglement generation or storing incoming quantum data at quantum repeaters, quantum routers, and user ends.


\textbf{(4) Quantum applications typically set stringent demands on the network, such as the desired end-to-end fidelity and request completion time.} Moreover, different quantum requests may have varying requirements for the network. In situations where the network fails to provide a specific service, numerous classical applications can still operate, albeit at a reduced quality. In contrast, quantum applications are far less tolerant; they cease to function if the network does not meet their stringent requirements. Thus, quantum networks should not simply adopt the ``best-effort'' approach of classical networks.

\textbf{(5) The selection of specific quantum error correction codes and methods should be determined based on the application's fidelity requirements and the resources and quality of the path.} Generally, longer paths and lower link quality require larger quantum error correction codes to achieve high fidelity. Additionally, the quantum devices such as quantum routers and quantum repeaters need to be prepared for quantum error correction.

%

If the quantum router/repeater is not sufficiently prepared in advance, the arrival of quantum data could lead to two potential scenarios. Firstly, the lack of available quantum resources to store the incoming quantum data on the router/repeater, may result in data loss. Secondly, an excessively prolonged wait for the quantum router/repeater to set up the necessary entangled state for quantum error correction could lead to decoherence of the quantum data.
If the quantum repeater is not prepared when the quantum data arrives, it will lead to packet loss or decoherence resulting from excessively long waiting times, making the quantum data unusable.

\section{System Design}
Given these characteristics of the quantum network, we can propose a new design approach. Specifically, our design encompasses the following aspects:

\subsection{Quantum Network Node Classification and Network Structure}
Quantum network nodes can be classified into three types: user ends, quantum routers, and quantum repeaters. The entire quantum network can be divided into two parts: the user-end network and the main network. The user-end network consists of user ends and quantum routers directly connected to these user ends (which can be referred to as edge quantum routers). This network is responsible for qubits transmission from the user end to its adjacent router, using the third-generation QR technology. 
The main network, comprising quantum repeaters and quantum routers, is responsible for transmitting qubits among the edge quantum routers or establishing entangled states between them. It is compatible with the first, second, and third-generation QR technologies. 

\begin{figure}
	\centering
	\includegraphics[width=\linewidth]{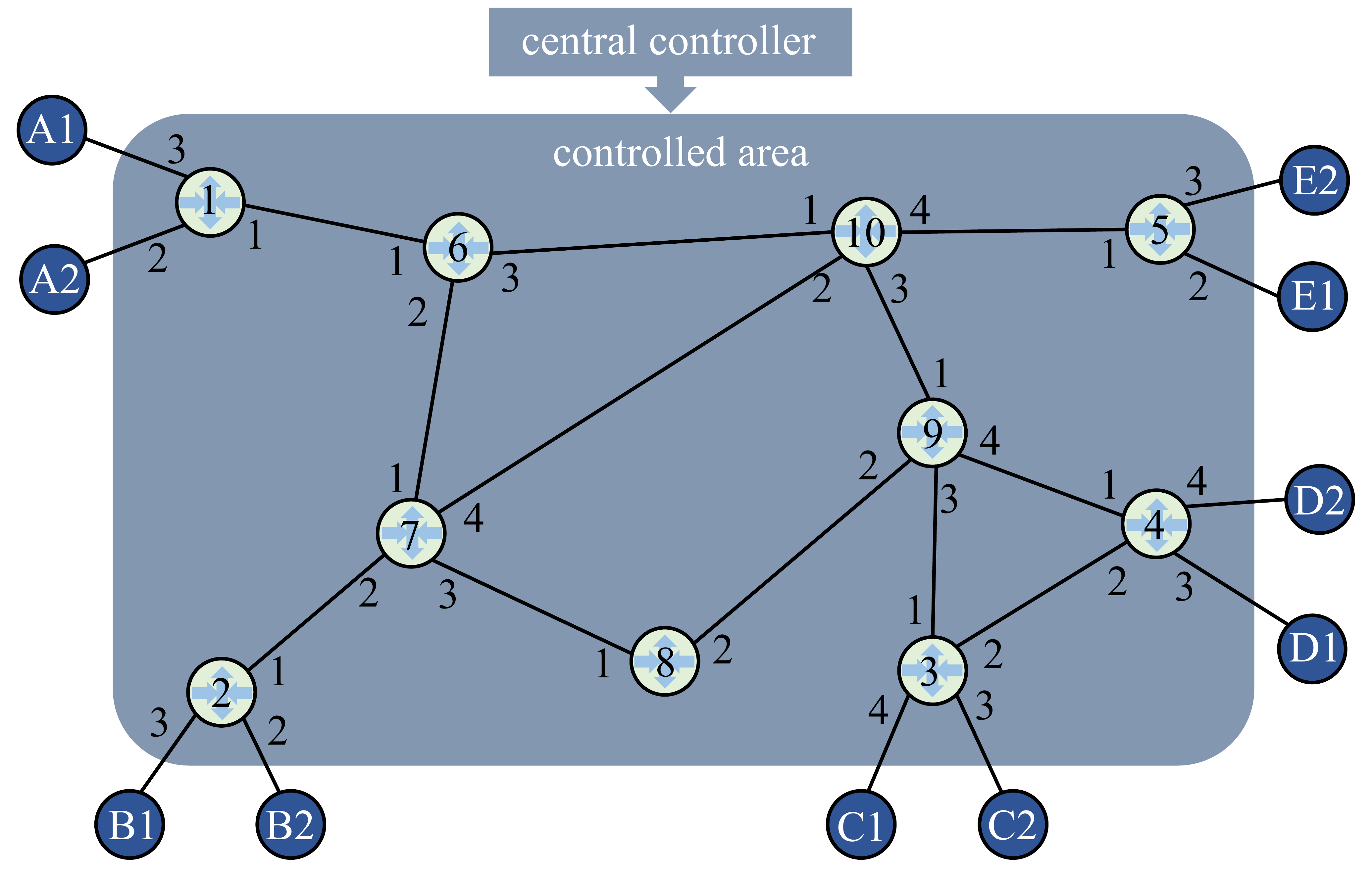}
	\caption{\label{fig:topo}Schematic of an Early Quantum Network.}
\end{figure}

As depicted in Fig.~\ref{fig:topo}, if a quantum request involves transmitting quantum data from User A.1 to User B.1, the main network's responsibility is to transfer the quantum data from Quantum Router 1 to Quantum Router 2. The segments from user A.1 to quantum router 1 and from quantum router 2 to user B.1 are handled by the user-end network.

Most existing research articles on quantum network routing operate under the assumption that all quantum nodes possess the capabilities to make requests and handle routing tasks, which means there's no clear distinction between user ends and quantum routers. Additionally, when determining routing paths, it's generally assumed that the network is aware of the quantum resource status of all nodes. However, in practical scenarios, user ends might have to spend quantum bit resources on local quantum tasks, and due to concerns about privacy, the user end might be reluctant to let the network have complete knowledge of its internal resource utilization. Furthermore, if the functions of user ends and quantum routers are integrated, it would place higher demands on the device's quantum resources. This is because their quantum resources would not only be required for local quantum tasks and joint tasks in the quantum network, but also for providing routing and repeater functions for remote quantum communication between other nodes.

Out of practical considerations and drawing an analogy from the classical internet, we separate the functions of user ends and quantum routers. Specifically, user ends do not possess the capabilities of quantum routers; they do not participate in constructing other requests. Conversely, quantum routers do not have the functionalities of user ends and do not make requests. This segregation of functions simplifies the network's operation and resource allocation, allowing for more efficient management.

The main network is designed to autonomously choose the appropriate quantum repeater technology based on different requests and the capabilities of quantum devices. For the first and second generation quantum repeater technologies that construct the quantum network using entangled states, the approach to network construction can be categorized into two based on the sequence of entangled state generation and the receipt of quantum requests: pre-construction and on-demand construction. Although the advantages and disadvantages of these two methods in various scenarios still warrant further exploration, our design mandates the use of the on-demand construction approach.
The on-demand method offers more flexibility. Depending on the necessity of constructing quantum states, it can utilize first and second-generation technologies when quantum state construction is required, or switch to third-generation QR technology when it is not.
Furthermore, this on-demand construction methodology empowers the central controller to dynamically adjust quantum resource allocation on network links in response to requests, thereby more effectively ensuring the realization of quantum requests, especially in situations where quantum resources are scarce in the network.

The user-end network consistently adopts the third-generation quantum repeater technology. Compared to the first and second generations of quantum repeater technologies, the third generation avoids complex coordinated operations with various quantum devices within the network. Moreover, user ends should possess capabilities to run multiple quantum applications, making them more advanced than the quantum routers and repeaters in the network. They are supposed to feature longer quantum storage times, enhanced precision in gate operations, and the capability to perform quantum error correction.. Hence, it's reasonable to assume that the performance of user-end devices is sufficient to support the use of the third-generation QR technology.

\subsection{Centralized Control and Dynamic Resource Allocation}
Eliminate autonomous system within the quantum network. The main network adopts centralized control with universal coordination across the entire network. A central controller issues rule sets and forwarding tables to quantum routers and repeaters.

Eliminating autonomous system sacrifices the right of different organizations or managers to independently manage parts of the main network, but this is necessary in the early stages of quantum network development. At this stage, quantum resources are extremely limited, and the duration of quantum storage is short. If different parts of the main network are controlled by different managers, it becomes challenging to realize quantum requests across regions.

The central controller designs customized plans for each quantum request based on its specific requirements and the real-time usage of network resources, adopting a connection-oriented approach with fixed paths and reserved resources.

Each quantum request has its distinct requirements, such as different end-to-end fidelities and request completion time. Additionally, the capabilities of the nodes within the network can vary; some nodes might only be able to use a specific generation of quantum repeater technology, while others may support multiple technologies. These factors necessitate that the central controller formulates customized plans for each quantum request.

Quantum memory in quantum devices is limited, making quantum networks more susceptible to congestion compared to classical networks if not properly managed. Therefore, to ensure the smooth operation of the network, it is essential to allocate resources efficiently at each quantum router, minimizing the risk of data loss caused by congestion. 

In the first and second-generation technologies, the simultaneous building approach, where nodes along the path concurrently start entanglement generation and swapping operations, necessitates collaboration among all nodes. To ensure the effectiveness of this collaboration, using a connection-oriented approach and reserving the necessary resources becomes vital. The central controller is responsible for issuing RuleSets to coordinate quantum devices on the path for collaborative operations. While the per-hop building approach supported by the second generation technology and the third generation technology can theoretically adopt a connectionless approach, not fixing the specific path of quantum frame transmission, the size of the quantum error-correcting codes in practical applications is influenced by the length and quality of the path. Therefore, a connection-oriented approach with a fixed path can simplify the determination of quantum error-correcting code size. Moreover, nodes on a fixed path can prepare in advance the resources needed for quantum error correction, thereby reducing transmission latency and the risk of packet loss. Specifically, the central controller needs to issue forwarding tables and notify quantum devices on the path in advance to prepare for quantum error correction. 

The connection-oriented and resource reservation mode allows the central controller to have more precise control over the utilization of quantum resources in various quantum devices within the main network. In many current quantum network routing schemes, the quantity of quantum resources on network nodes can impact the final path selection. Consequently, such a scheme enables the central controller to choose the current optimal path for each service.


User ends are purposefully excluded from the management of the central controller. This exclusion aims to diminish the processing complexity faced by the central controller by reducing the number of quantum devices under its direct control. Furthermore, this approach serves to protect the privacy of user ends.

It's important to note that edge quantum routers (connected to user ends) possess the authority to independently handle local quantum requests. This means they have the right to decide how to utilize their quantum resources. Consequently, the central controller lacks awareness of their quantum resource usage and must inquire whether they have sufficient quantum resources to fulfill a request.

As the quantum network matures into a large-scale system with abundant quantum resources and enhanced quantum storage capabilities, the existing design can be suitably modified. For example, autonomous systems can be reintegrated into the main network, each domain being under its own administration. Broad routing planning and decomposition of indicators could be performed on a master controller, delegating detailed management and implementation tasks to various domain-specific controllers.

\subsection{Design of Quantum Frame}
 In the transmission of qubits, the header of the quantum frame contains only the request ID and path ID, without destination addresses, port numbers, and other information. A quantum frame consists of a header composed of classical bits, a payload made up of qubits, and a trailer also made up of classical bits.

In classical networks, session differentiation primarily relies on the IP five-tuple. However, this method often depends on the destination address for routing decisions, lacking sufficient granularity. In contrast, we have adopted a more customized approach in quantum network, tailoring to the specific needs of each request. Even for two quantum requests with the same source and destination addresses, different transmission paths might be chosen due to their different requirements and the distinct real-time availability of quantum resources in the network.

During the setup of connection establishment, the central controller provides key details to the quantum repeaters/routers along the selected path, including the request ID and path ID corresponding to each quantum request, as well as the port identifiers on the respective quantum repeaters/routers. This method means that in the qubits transmission process, the header of the quantum frame does not need to carry information such as the destination address or port number. Instead, it uses the request ID and path ID for identification and routing, ensuring that the data is accurately transmitted to the intended recipient.

%

We stipulate that the size of quantum frames within the same request should be uniform.

Given the limited quantum memory in quantum devices, the size of quantum frames within the same request should be identical. This uniformity aids in facilitating resource reservation on quantum routers and repeaters, as well as preparing the necessary resources for quantum error correction. If the sizes of quantum data frames within the same request were to vary, it would be challenging for the quantum nodes to predict the size of upcoming frames, thus complicating the preparation of sufficient quantum resources in advance.

%

\subsection{Design of Quantum Devices}
Network interface components and their buffers are eliminated in quantum user ends, adopting a strategy of direct transmission and immediate processing for qubits.

Due to the fragile nature of quantum states, if qubits arriving at the user end need to be stored in a buffer before being transferred to the local quantum memory, this transfer process can lead to a degradation in the quality of the quantum state. Direct transmission to quantum memory can avoid this issue. Moreover, removing separate network interface components enables the central processor to schedule qubits more efficiently. Besides, considering the complexity of quantum state preparation and the scarcity of quantum memory, it is more resource-efficient to consolidate quantum memory within the computer itself, rather than using it in network interface buffers.

Of course, if quantum user ends develop to a stage comparable to the maturity of classical computers, with abundant quantum memory and highly precise quantum gate operations, this distinction can be reinstated.

In addition, quantum routers internally employ a shared memory model for quantum storage, without dedicating memory to specific interfaces. Resource allocation is managed by the central controller as needed.
\section{identifier}
In this chapter, we introduce the network identifiers required in quantum networks.

\textbf{(1) User End Identifier.} User End Identifier, also known as the hostname. Hosts under the same quantum router share a unique hostname prefix, which must be distinct network-wide. While the central controller isn't required to track individual hostnames, it must recognize the quantum router associated with each prefix. This system permits the addition or removal of user ends without informing the central controller.

\textbf{(2) Quantum Router/Repeater Identifier.} Quantum routers/repeaters are uniquely coded throughout the entire network, with each having a fixed address identifier. It is reasonable to assume that quantum routers and repeaters are generally stationary. If the devices are relocated, added, or removed, the central controller must be informed promptly.

\textbf{(3) Quantum Router/Repeater Port Identifier.} The port number serves as a local identifier; however, the central controller is aware of the identifiers for all quantum routers/repeaters throughout the network and their respective port identifiers, as well as the specific quantum routers or repeaters connected to each port. For quantum routers connected to user ends, the central controller does not need to know which specific port a user is connected to.

\textbf{(4) Quantum Request Identifier.} The quantum request identifier includes the names of both the requesting and target user ends, their respective quantum applications, as well as the requirements for the quantum network. These requirements may include the amount of quantum data to be transmitted, the desired expression of the entangled state to be formed, and the minimum acceptable end-to-end fidelity and request completion time, among others.

When a connection is successfully established, the central controller assigns a globally unique request ID to the quantum request. For local quantum requests, the connected quantum router assigns a locally unique request ID. Once the quantum request is completed, the request ID can be reallocated for use by other requests.

\textbf{(5) Path Identifier.} Considering the relatively small scale of early-stage quantum networks, there are not many paths in the network, so each path can be assigned a globally unique identifier. For paths that are frequently used within the network, fixed path identifiers can be allocated; for less commonly used paths, the central controller can dynamically assign a temporary path identifier and inform the relevant quantum routers. These temporary identifiers can be revoked after their use has concluded.

In quantum networks, a single request may involve the use of multiple paths. For instance, in two-end communication, multiple paths can be selected to increase throughput. Another scenario is multi-end communication, such as distributing multipartite entanglement, where multiple paths are necessary to fulfill the request. During the transmission of qubits, the packet header needs to include the request ID and the path ID.

During the connection establishment process, the central controller informs the quantum router of the port numbers associated with each path ID. 


\textbf{(6) Qubit Identifier.} A qubit is identified not only by its physical address but also by state identifiers, such as unused or used. When a qubit is in an entangled state, additional identifiers are required for the entanglement, including an ID, generation time, expiration time, Bell state index, associated node IDs, and its current  status, which may be alive, measured, or discarded. For qubits allocated to a request, the identifier also needs to carry the request ID.

\section{the specific process of quantum request}
The completion of a quantum request consists of two steps: the establishment of a connection, and the transmission of quantum data or the establishment of remote entangled states. To facilitate a better understanding for readers, we specifically introduce three instances here: a local quantum request between two parties, a remote quantum request between two parties, and a remote quantum request involving three parties.

\begin{figure}
	\centering
	\includegraphics[width=\linewidth]{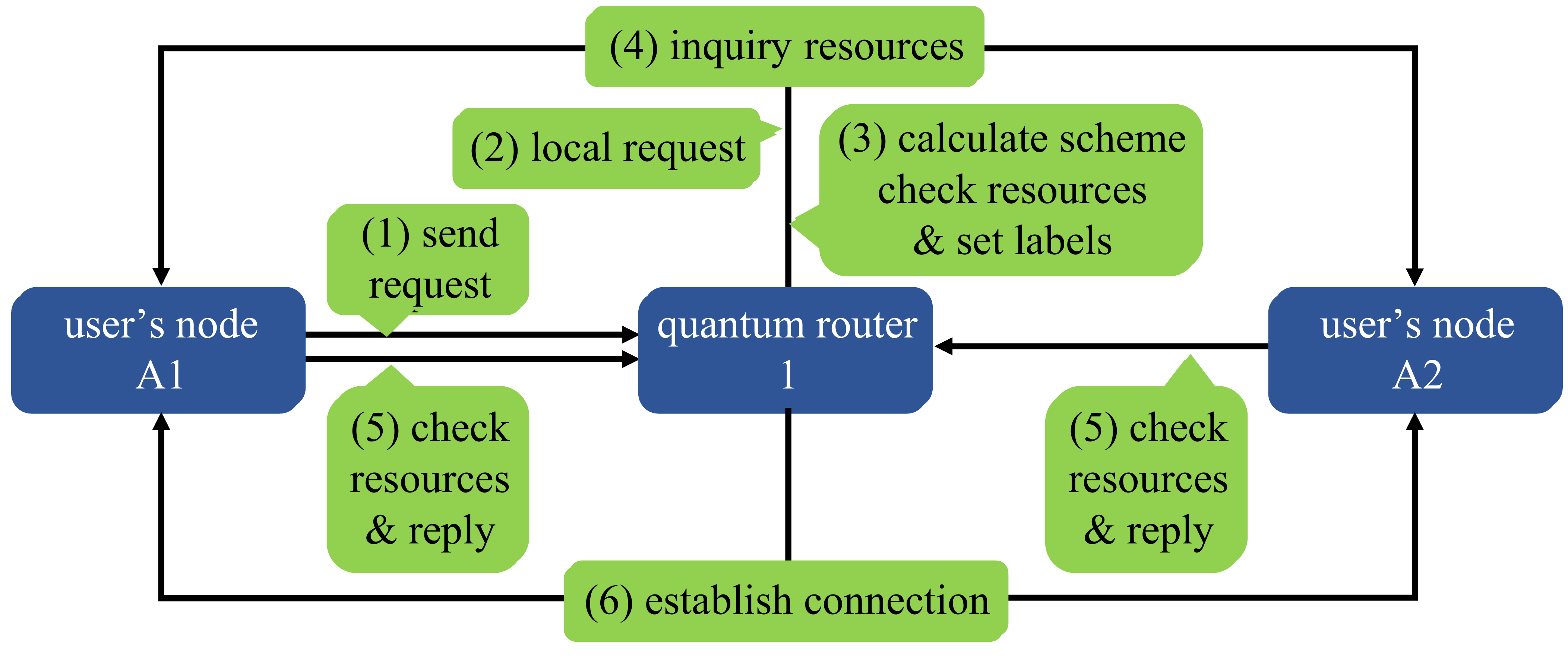}
	\caption{\label{fig:local}Flowchart of local quantum request implementation.}
\end{figure}

\textbf{Example 1: User end A.1 wishes to engage in quantum communication with user end A.2.}
Since A.1 and A.2 are connected to the same quantum router, this quantum request is considered a local quantum request. Fig.~\ref{fig:local} shows the flowchart of the process for implementing the request.

\textbf{Connection establishment:}

Step 1, user end A.1 sends a request to quantum router 1. The content of the request includes the name of the quantum application initiating the request, the name of the target user end A.2 and the name of the quantum application receiving the request on A.2, as well as the expected standards for the quantum network. These requirements could encompass the quantity of quantum data to be transmitted, the minimum acceptable end-to-end fidelity, throughput, latency, and so on.

Step 2: Upon receiving the request, quantum router 1 verifies that the target user end A.2 is a local user. Based on the content of the request and information such as the known link quality, it calculates the necessary quantum encoding method, quantum error correction technique, the size of the quantum data frames, the transmission rate, and so on. If so, it assigns a local request ID tag to this request, records the corresponding information and port details, and communicates the request, the calculated scheme, and the ID tag to both user ends A.1 and A.2.

Step3, user ends A.1 and A.2, upon receiving the message, examine whether their quantum memory resources can meet the calculated scheme of quantum router 1, and then send the results back to quantum router 1. 

Step4, Based on the responses from user ends A.1 and A.2, quantum router 1 either informs both user ends that the connection for their request has been successfully established, or notifies them that the request cannot be processed at the moment.

\textbf{Quantum data transmission:}

User ends A.1 and A.2 allocate the required quantum memory for the request. Subsequently, user end A.1, using the third-generation quantum repeater technology as the requirements, marks the request ID in the packet header and sends the quantum frame to quantum router 1 for quantum error correction. Finally, the quantum frame is forwarded to user end A.2. 

\textbf{Example 2: User end A.1 wishes to engage in distributed quantum computing with user end C.1.}

\begin{figure*}
	\centering
	\includegraphics[width=0.9\linewidth]{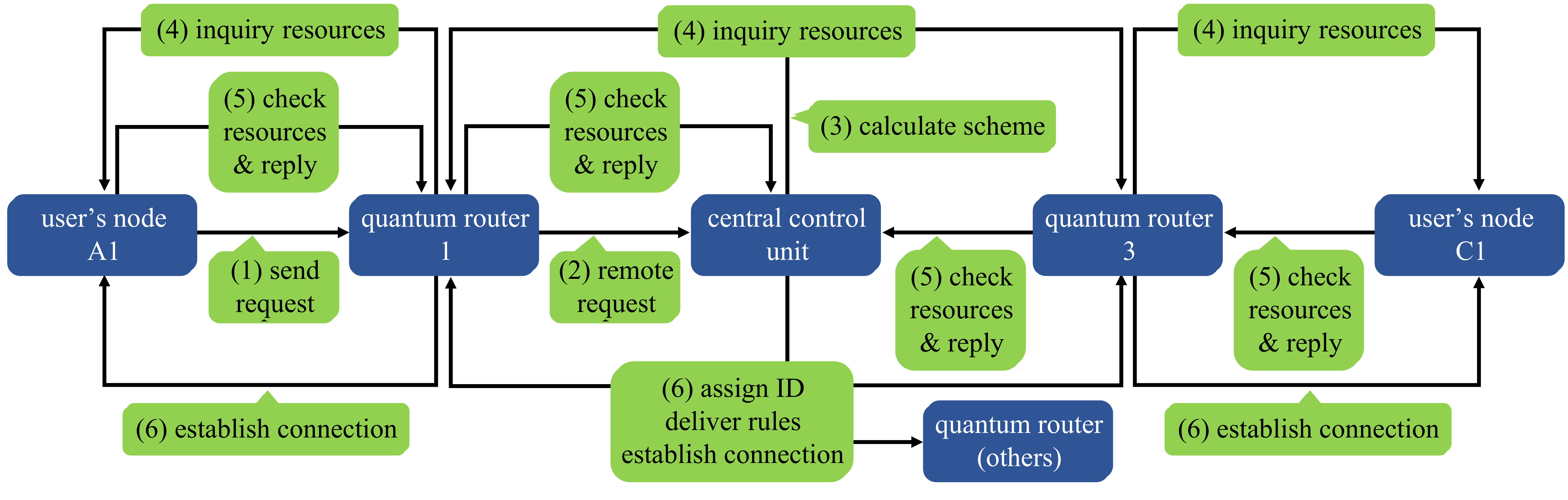}
	\caption{\label{fig:remote}Flowchart of remote quantum request implementation..}
\end{figure*}

Since A.1 and C.1 are connected to different quantum routers, this quantum request is classified as a remote request. Fig.~\ref{fig:remote} shows the flowchart of the process for implementing the request.

\textbf{Connection establishment}:

Step 1. User end A.1 sends a quantum request to quantum router 1.

Step 2. Quantum router 1, determining that the target user end C.1 is not a local user, forwards the quantum request to the central controller.

Step 3. The central controller, based on the requirements of the request and the usage of quantum resources in the main network, as well as information on link quality, makes a decision. It inquires whether quantum router 1, quantum router 3, and user ends A.1 and C.1 have sufficient quantum resources to implement the plan.

Step 4. Quantum routers 1 and 3, along with user ends A.1 and C.1, check their quantum memory resources to ensure they can meet the request, and respond to the central controller.

Step 5. Upon receiving affirmative responses, the central controller allocates a path ID and a globally unique request ID for the request. It then dispatches the IDs along with corresponding forwarding tables or rule sets to the quantum routers and repeaters on the selected path and informs both user ends that the connection for their request has been successfully established. If the central controller receives negative responses, it informs user ends A.1 and C.1 that the request cannot currently be processed.

\textbf{Quantum data transmission:}

\begin{figure}
	\centering
	\includegraphics[width=\linewidth]{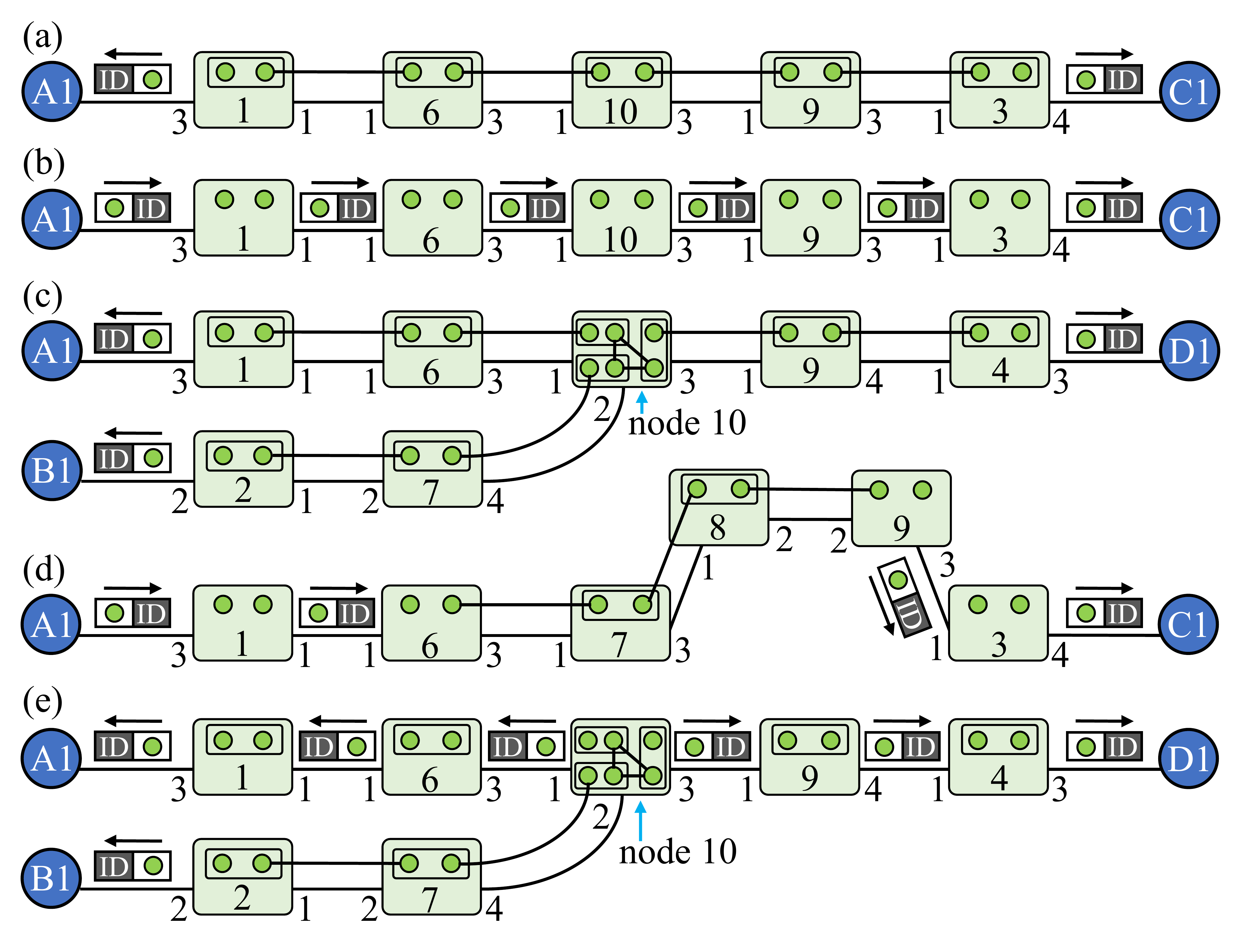}
	\caption{\label{fig:request}Schematic of quantum request implementation: (a, b, d for example 2. (c, e) for example 3.}
\end{figure}

Since the transmission in this case is conducted through the main network, a variety of technologies can be employed. Here, we first demonstrate how to transmit using second-generation quantum repeater technology (see Fig.~\ref{fig:request}(a)).

First, encoded entangled states are generated between quantum routers and quantum repeaters along the designated path. Then, quantum routers/repeaters in the middle of the path perform entanglement swapping operations on the qubits at their specified ports. This process establishes long-distance entangled states between the quantum routers connected to the user ends. After applying quantum error correction to the established remote entangled states, quantum routers 1 and 3 forward them to user nodes A1 and C1.  It is important to note that one entangled state can be used to transmit one qubit. Therefore, to transfer multiple qubits, multiple remote entangled states need to be established.

Now, let's assume that the main network employs third-generation quantum repeater technology for this request (see Fig.~\ref{fig:request}(b)). Let's see how it works.

Similar to the previous example, user end A.1 prepares the quantum frame as required and sends it to quantum router 1. Quantum router 1 then performs quantum error correction based on the request ID and forwards it to the designated port for transmission to the next hop. In this manner, the frame is hop-by-hop forwarded along the specified path, undergoing quantum error correction, until it reaches user end C.1.

\textbf{Example 3: User end A.1 wishes to engage in quantum sensing with user ends B.2 and D.1., requiring the establishment of GHZ state among the three parties.}

For this type of request, there are several possible approaches. In this instance, we opt for a method that begins with selecting a quantum router or quantum repeater within the network to serve as a factory node. This factory node is tasked with preparing the required tripartite entangled state and subsequently distributing each segment of this entangled state to the respective user ends. Key routing decisions in this approach include determining which quantum device will act as the factory node and selecting the optimal paths from this factory node to each of the user ends.

\textbf{Connection establishment:}

Step 1: User end A.1 sends a request to quantum router 1.

Step 2: Quantum router 1 forwards the request to the central controller.

Step 3: The central controller formulates a plan and inquires whether quantum routers 1, 2, 4, and user ends A.1, B.2, D.1 have the resources and conditions to fulfill the plan.

Step 4: Each quantum router and user end assesses their own resources and conditions to determine if they can meet the plan's requirements, and then they respond to the central controller with their findings.

Step 5: Upon confirming that all relevant devices can meet the requirements, the central controller assigns a request ID and path ID. It then informs all involved quantum devices of the operations they need to perform. For this instance, quantum router 10 is selected as the factory node, the three marked paths are the paths from user ends to factory node selected by the central controller.

\textbf{Remote Entangled States establishment}:

The main network can be implemented using second-generation quantum technology(see Fig.~\ref{fig:request}(c)). Initially, remote entangled states are generated along three paths: 1-6-10, 2-7-10, and 4-9-10, thereby establishing entanglement between routers (1,10), (2,10), and (4,10). Once these remote entangled states are established, quantum router 10 prepares a local GHZ state. A Bell-state measurement (BSM) is then performed between this local GHZ state and the three remote entangled states. The measurement results are forwarded to terminal quantum routers 1, 2, and 4, which apply the necessary corrections to their local qubits based on the received information. The corrected qubits are subsequently forwarded to the user nodes, ultimately forming a three-party GHZ state shared among the end users.

Thus far, we have considered a scenario in which the core network employs a uniform quantum repeater technology for path construction. However, a hybrid approach integrating multiple generations of quantum technologies can also be adopted. As illustrated in Fig.~\ref{fig:request}(d), in example 2, the central controller selects the path 1-6-7-8-9-3, where the links 1-6 and 9-3 utilize third-generation technology, while the intermediate section (6-7-8-9) employs second-generation technology. Similarly, in Example 3 (see Fig.~\ref{fig:request}(e)), the segments 1-6-10 and 4-9-10 employ third-generation technology, whereas 2-7-10 relies on second-generation technology. This hybrid deployment strategy enables more flexible and efficient utilization of existing quantum networking technologies.

\section{Simulation}
In this section, we simulated our designed quantum network model. The simulation was conducted using the discrete-event quantum network simulation platform, NetSquid. NetSquid enables the modeling of various physical components, such as quantum sources, detectors, memories, processors, and quantum/classical channels, while also simulating noise and loss. We adopted the hardware models for a platform-agnostic abstract model as described in \cite{Avis_requirements_2023}. This choice was made to avoid constraining early quantum networks to a specific platform. This model considers depolarizing noise in all gates and photon emission, as well as amplitude-damping and phase-damping noise in quantum memories.


Compared to the previous version\cite{Avis_requirements_2023}, our code has grown by approximately 7,000 lines to implement the following features:

1. Simulations were conducted on the network topology shown in Fig.~\ref{fig:topo}, covering both the main network and user-side networks. This includes devices such as the central controller, user nodes, and terminal repeaters. The request process was also implemented.

2. A routing function was added, utilizing the Dijkstra algorithm to determine paths within the network.

3. The assumption of hardware limitations requiring repeaters to operate sequentially was removed. Instead, it is assumed that each link in the network can perform multiple operations—such as entanglement generation, swapping, discarding, and regeneration—in parallel. Additionally, nodes are assumed to support multiple qubits.

4. Implemented a multi-node cutoff and discard mechanism. This allows simulation of scenarios where some links in the path have completed entanglement swapping, but the entangled states are actively discarded by the network due to the unavailability of other links.

5. Simulated not only requests for generating bipartite entangled states but also requests for generating multipartite entangled states, such as GHZ states\cite{avis_analysis_2023}.

6. To enable operation in complex quantum network topologies rather than being restricted to repeater chains, the active and passive modes for negotiating entanglement generation between nodes were changed from being bound to nodes to being bound to node ports. This allows a single node to have different modes across its ports, whereas previously, each node supported only one mode.

7. During entanglement swapping, priority is given to entangled states generated earlier. If the generation times are identical, entangled states are sorted based on spatial coordinates. An analysis of the success probability for entanglement swapping was also added.

8. To make full use of network resources, a regeneration process is immediately initiated when a qubit at a node is discarded or entanglement swapping is completed. This ensures continued service for the request until it is fulfilled.

9. Added an analysis of how the central controller's allocation of network resources affects the fidelity of the final entangled states and the request completion time.

The formula for calculating fidelity $F$ is as follows:
\begin{equation}
	F:=\left(t r \sqrt{p^{1 / 2} \sigma \rho^{1 / 2}}\right)^2
\end{equation}
where $\rho$ represents the density matrix of the entangled state when it arrives at the user's end, while $\sigma$ represents the density matrix of the ideal entangled state.

The noise models used in the simulation include:
The fibre loss model was used. The probability of qubit loss is:
\begin{equation}
	p_{\mathrm{loss}} = 1 - 10^{- L/2 p_{\mathrm{L}}  / 10}
\end{equation}

where $L$ represents the distance between quantum routers, while $p_L$ represents the photon survial probability per channel.

The built-in T1T2NoiseModel from NetSquid was used for quantum memory noise. Specifically, the decay of the state stored in a communication qubit or storage qubit is modeled using a noise model based on the relaxation time ($T_1$) and dephasing time ($T_2$).

Entanglement swapping is modeled using the depolarize noise model, characterized by the parameter $s_q$. $s_q=1$ indicates the absence of noise.

The probability $p$ of successfully generating the entangled state on the link and the initial density matrix $\rho$ are provided by the following equations. 

\begin{equation}
	\begin{aligned}
		&p=  p_{\mathrm{T}}+p_{\mathrm{F} 1}+p_{\mathrm{F} 2}+p_{\mathrm{F} 3}+p_{\mathrm{F} 4}, \\
		&\rho=  p_{\mathrm{T}}\left|\Psi^{ \pm}\right\rangle\left\langle\Psi^{ \pm}\right|+p_{\mathrm{F} 1} \frac{|01\rangle\langle 01|+|10\rangle\langle 10|}{2}+ p_{\mathrm{F} 2} \frac{|00\rangle\langle 00|+|11\rangle\langle 11|}{2},\\
		&p_{\mathrm{T}}= \frac{1}{2} p_{\mathrm{A}} p_{\mathrm{B}} V\left(1-p_{\mathrm{dc}}\right)^2\\
		&p_{\mathrm{F} 1}=\frac{1}{2} p_{\mathrm{A}} p_{\mathrm{B}}(1-V)\left(1-p_{\mathrm{dc}}\right)^2\\
		&p_{\mathrm{F} 2}= \frac{1}{2} p_{\mathrm{A}} p_{\mathrm{B}}(1+V) p_{\mathrm{dc}}\left(1-p_{\mathrm{dc}}\right)^2\\
		&p_{\mathrm{F} 3}= 2\left[p_{\mathrm{A}}\left(1-p_{\mathrm{B}}\right)+\left(1-p_{\mathrm{A}}\right) p_{\mathrm{B}}\right] p_{\mathrm{dc}}\left(1-p_{\mathrm{dc}}\right)^2 \\
		&p_{\mathrm{F} 4}=4\left(1-p_{\mathrm{A}}\right)\left(1-p_{\mathrm{B}}\right) p_{\mathrm{dc}}^2\left(1-p_{\mathrm{dc}}\right)^2 .\\
		&p_{\mathrm{A}}=p_{\mathrm{B}}=p_{\mathrm{loss}}p_{\mathrm{de}}\\
		&\left|\Psi^{ \pm}\right\rangle=\frac{1}{\sqrt{2}}(|01\rangle \pm|10\rangle)\\
	\end{aligned}
\end{equation}
where $V$ represents visibility of photon interference, while $d_e$ represents photon detection probability excluding attenuation losses. Parameters such as $p_L$, $L$, and $d_e$ significantly influence the success probability, while $V$ exerts a substantial influence on the density matrix.


In our simulation, we assumed that entanglement generation between nodes follows the double-click scheme within the MeetInTheMiddle method\cite{li_survey_2024} and that quantum routers are equipped with quantum memory.
Certain simplifications were made in the simulation. For instance, the transmission of qubits from user ends to edge quantum routers did not involve actual quantum encoding. Instead, it was assumed that the photon loss probability in this segment was zero. Additionally, the distance from each node to the central controller was set to 100 km, the distance from each user end to its edge quantum router was set to 1 km, and the distances between nodes within the main network were assumed to be equal.

Entanglement generation was performed on-demand. When the central controller received a request, it allocated a specific amount of network resources for that request. For example, if the request required 100 pairs of entangled states, the network allocated 10 pairs of entangled states per link along the path. These resources were reused repeatedly until 100 end-to-end entangled pairs were successfully generated. Table \ref{table:data} shows the best achievable values of certain parameters in quantum networks based on current experiments, as well as the values used in each figure presented in our simulations. 

\begin{table*}[htbp]
	\caption{\label{table:data} Quantum network parameters: best experimental values and simulation values used in each figure. $V$: visibility of photon interference, $s_q$: swap quality, $L$: distance between quantum routers, $p_L$: photon survial probability per channel, $s_p$: swap probability, $d_e$: photon detection probability excluding attenuation losses, $T_1$: relaxation time, $T_2$: dephasing time, $N$: number of network resources allocated to the request by the central controller.}
	\centering
	\begin{tabular}{|c|c|c|c|c|c|c|c|c|c|c|}
		\hline
		& $V$ & $s_q$ & $L$(km)&$p_L$(dB/km) & $s_p$ & $d_e$  &$T_1$(h) &$T_2$(s)&cutoff time(s)&$N$\\
		\hline
		Experiment\cite{Avis_requirements_2023} & 0.9 & 0.83 &- & 0.2 &[0,1] &$5.1\times 10^{-4}$ &10&1&-&-\\
		\hline
		Fig.5(a1)(a2) &[0.9,1]  & [0.8,1] &50 & 0.2 &1 &1 &10&1&-&2\\
		\hline
		Fig.5(b1)(b2) &  1& 1&[10,190] &[0,0.3] &1 &1&10&1&-&2\\
		\hline
		Fig.5(c1)(c2) &  1& 1&50 &0.2 &[0.5,1] &[0.1,1]&10&1&-&2\\
		\hline
		Fig.5(d1)(d2) &1 &1 &[10,190] &0.2 &1 &1&10&[1,100]&-&2\\
		\hline
		Fig.6 & 1 &1 &50,100 &0.2 &1 &0.2,0.5,1&10&1&[0.1,1]&2\\
		\hline
		Fig.7 &1 &1&50 &0.2 &1 &0.2,0.5,0.7,1&10&1&-&[1,101]\\
		\hline
		Fig.8 & 1&1& 50&0.2 &1 &0.2&10&1&-&1,6,11,16\\
		\hline
	\end{tabular}
\end{table*}

We attempted to use parameters from current experiments as much as possible. However, if all parameters were set to their present experimental values, the probability of entanglement generation for a basic link with $L=50$ km would be as low as $10^{-8}$. This would result in poor fidelity for the final entangled states and a completion time exceeding the simulation's feasible range. Therefore, we adjusted the range of certain parameters to anticipate the capabilities of early quantum networks.

For each set of parameters, 100 trials were conducted, and the average results are presented.

For example 2, assume that user A1 issues a request to establish two pairs of end-to-end entangled states with C1. The central controller determines the route in the main network as Router 1-6-10-9-3 and allocates two pairs of entangled state resources for each link in this route. 

\begin{figure*}[htbp]
	\centering
		\centering		
		\includegraphics[width=\linewidth]{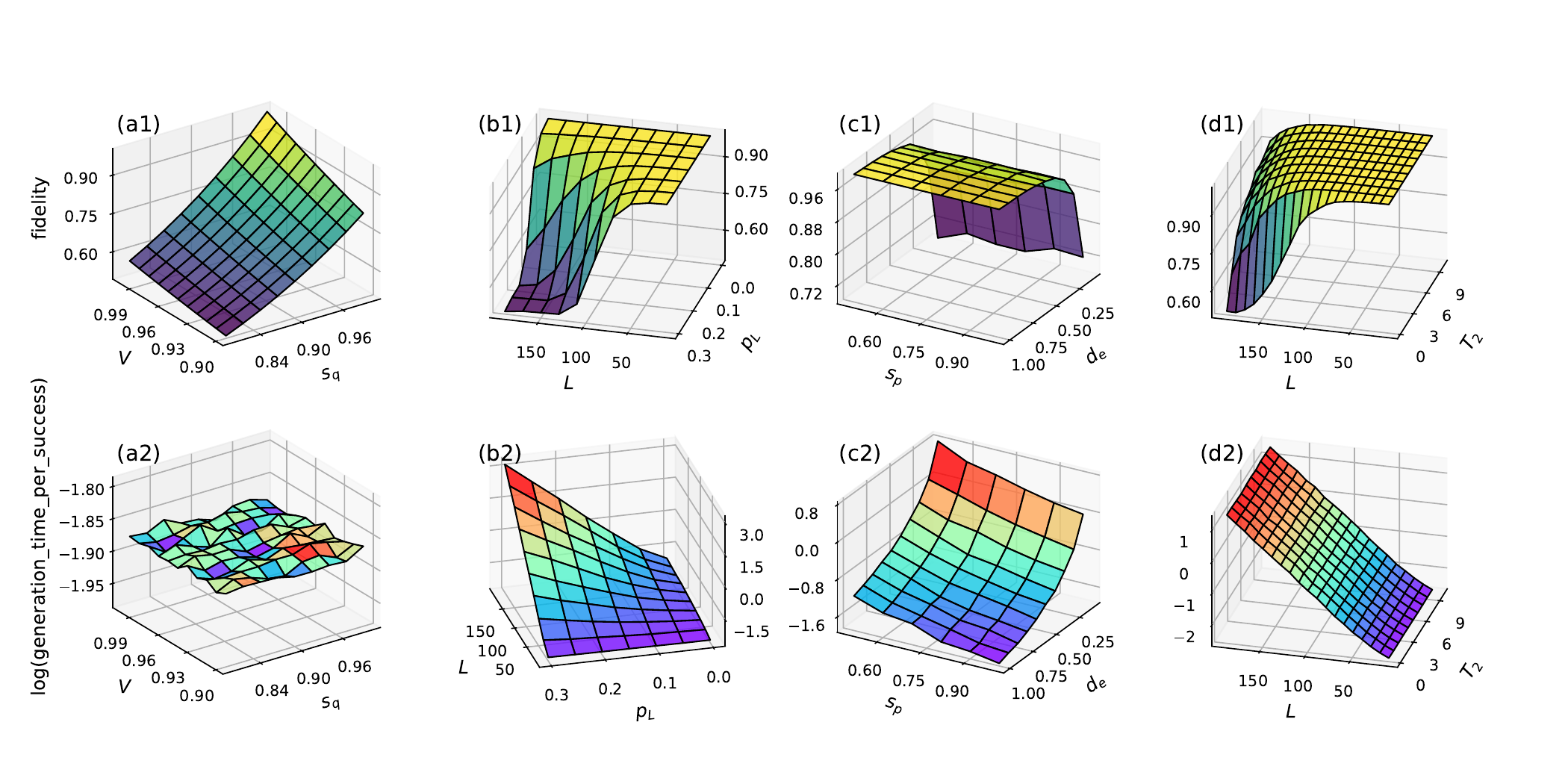}
		\caption{\label{fig:data1}The impact of parameters $V$, $s_q$, $L$, $p_L$, $s_p$, $d_e$, $T_2$ on the end-to-end entangled state fidelity and request completion time.}
\end{figure*}

Fig.~\ref{fig:data1}(a1) and (a2) illustrate the impact of visibility and swap quality on the fidelity of the end-to-end entangled state. 

As shown in the figure, the better the performance of visibility and swap quality, the higher the fidelity of the resulting entangled state. Within the range of simulated data, improvements in swap quality have a slightly greater effect on fidelity compared to visibility. These two parameters do not affect the entanglement generation time. The fluctuations shown in Fig.~\ref{fig:data1}(a2) are due to randomness and remain within the error margin, indicating statistically consistent.



Fig.~\ref{fig:data1}(b1) and (b2) illustrate the impact of $p_L$ and L on the fidelity of the end-to-end entangled state and the request completion time. Notably, an increase in L not only extends the classical communication time, causing quantum bits to experience longer decoherence processes, but also raises the probability of photon loss, thereby reducing the success probability of entanglement generation.

From the figure, it can be observed that smaller values of p loss length and L result in higher fidelity of the end-to-end entangled state and shorter request completion times. Taking the current optical fiber with $p_L = 0.2$ as an example, fidelity begins to decrease significantly when L exceeds 90 km.
As for the request completion time, when $L > 130 km and p_L > 0.2$, the time increases significantly, making it difficult to discern the effects of L and $p_L$ in other ranges. Therefore, a logarithmic scale is used to represent the request completion time in this context.


Fig.~\ref{fig:data1}(c1) and (c2) demonstrate the effects of detector efficiency and swap probability on the fidelity of the end-to-end entangled state and the request completion time.

For the fidelity, detector efficiency has a greater impact, while swap probability has a relatively smaller effect. This is because even if entanglement swapping fails, the only consequence is the discard of the entire chain's entangled state, followed by regeneration. This does not affect the survival time of the final entangled state but only influences the request completion time.
In this figure, the minimum value of detector efficiency is set to 0.1, whereas the current experimental value is approximately $5.1e^{-4}$, suggesting that the actual performance could be significantly worse than shown in the figure.

Fig.~\ref{fig:data1}(d1) and (d2) illustrate the impact of $T_2$ and $L$ on the final results. When $T_2 = 1s$, the fidelity shows a significant decline as the inter-node distance exceeds 100 km. Additionally, Fig.~\ref{fig:data1}(d2) indicates that $T_2$ has no effect on the request completion time.

%
%

For example 3, assume the initial fidelity of the locally generated GHZ state at node 10 is 0.98. A correlated depolarization noise model is applied to this local GHZ state. 

%

%


\begin{figure}[htbp]
	\centering
	\centering		
	\includegraphics[width=\linewidth]{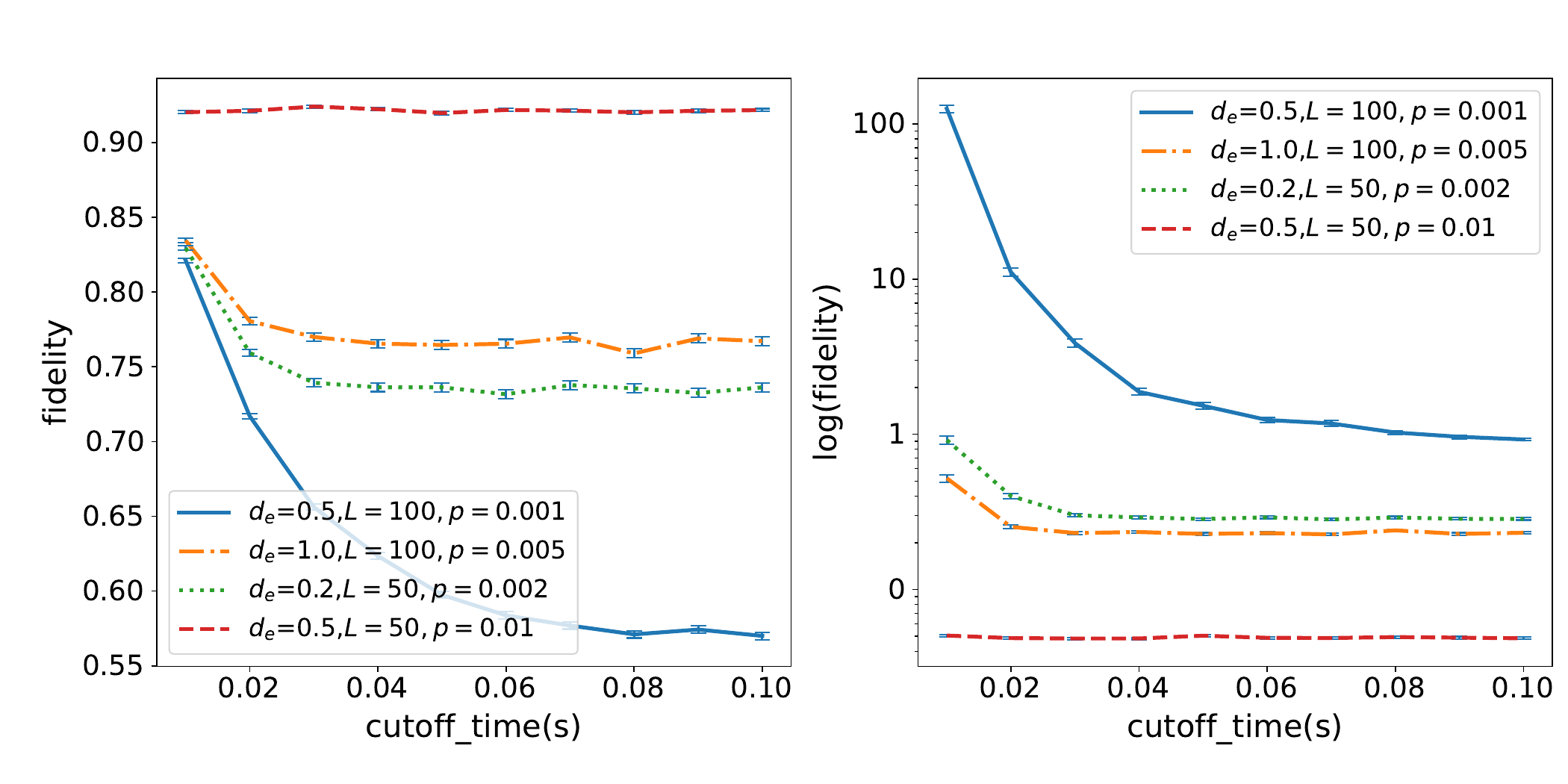}
	\caption{\label{fig:data2}The impact of cutoff time on the end-to-end entangled state fidelity and request completion time.}
\end{figure}

Fig.~\ref{fig:data2} investigates the impact of cutoff time on the final results. Cutoff time refers to the threshold beyond which an entangled state is actively discarded by the network to ensure that its fidelity meets the required standard. A truncation time of 0 indicates that both links on either side of a node must generate entanglement simultaneously for entanglement swapping to occur; otherwise, the entangled state is discarded. This corresponds to a scenario where quantum routers lack quantum storage capabilities.


From the figure, it can be observed that as the truncation time decreases, the fidelity of the final entangled state indeed improves. However, this comes at the cost of an increase in request completion time, as more entangled states are discarded by the network. Therefore, there is a trade-off between fidelity and request completion time in this scenario.

Furthermore, by comparing the data in the figure, we observe that as the inter-node distance increases and the entanglement generation probability decreases, the impact of reducing the cutoff time becomes more pronounced.


\begin{figure}[htbp]
	\centering
	\centering		
	\includegraphics[width=\linewidth]{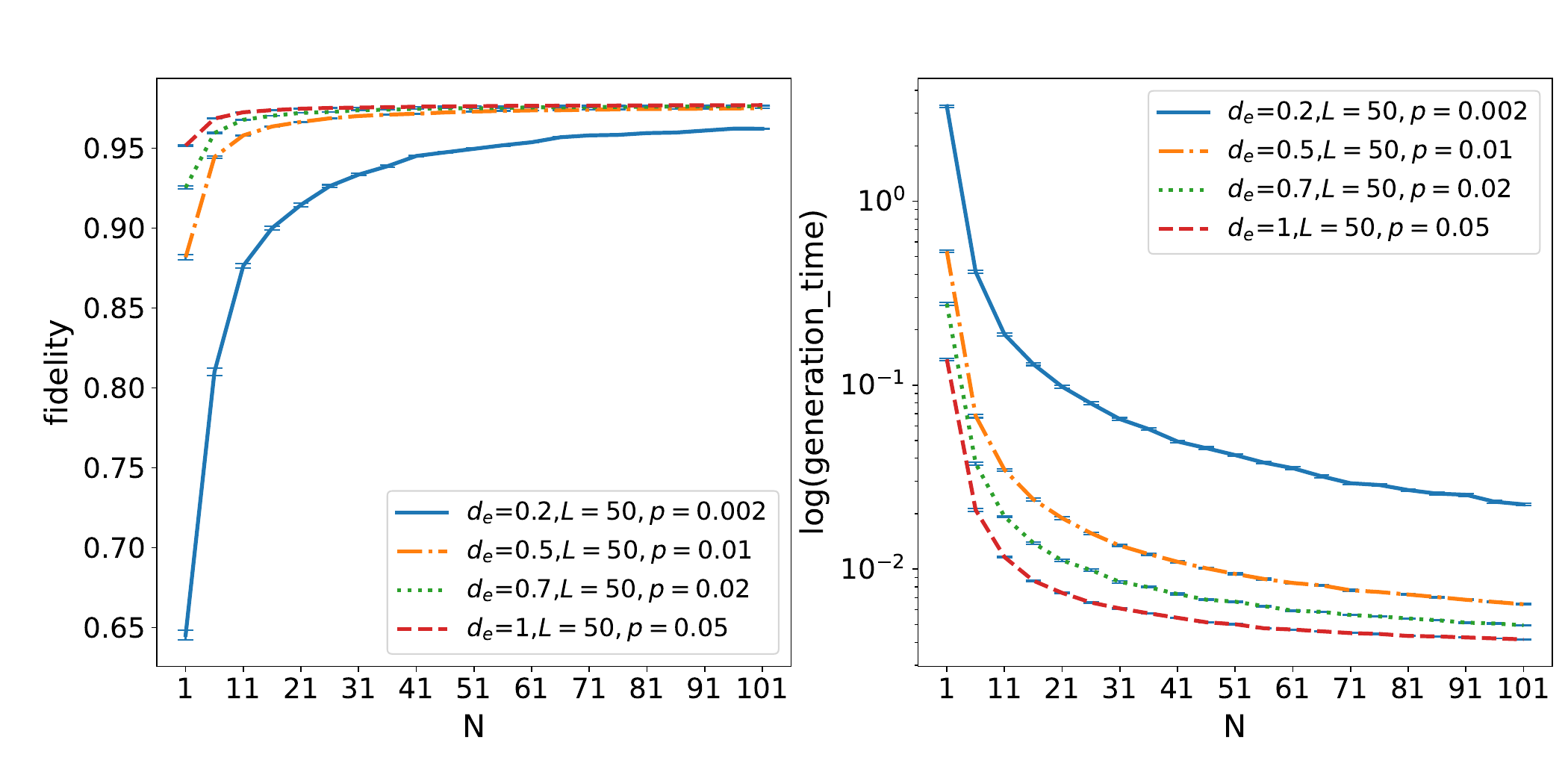}
	\caption{\label{fig:data3}The impact of number of network resources allocated to the request by the central controller on the end-to-end entangled state fidelity and request completion time.}
\end{figure}

Fig.~\ref{fig:data3} examines the impact of allocating different resources by the central controller for a request to generate 10 GHZ states on the final entangled state's fidelity and the request completion time. For instance, $N = 5$ means that each router port involved in the path construction provides 5 qubits and attempts to perform entanglement generation operations. Qubits that successfully form remote GHZ states proceed to the next round of entanglement generation, swapping, and other operations until 10 GHZ states are completed.  

Given the low probability of entanglement generation, allocating more resources than required (i.e., N > 10) can improve the fidelity of the final entangled state and reduce the request completion time.  Similar to Fig.~\ref{fig:data2}, when the entanglement generation probability decreases, the effect of increasing resource allocation becomes more significant.

Since the quantum request process we designed begins entanglement generation only after quantum routers and the central controller complete classical communication negotiations, this portion of classical communication time does not affect the fidelity of the final entangled state. Instead, it appears as a constant within the request completion time.


\begin{figure}[htbp]
	\centering
	\centering		
	\includegraphics[width=\linewidth]{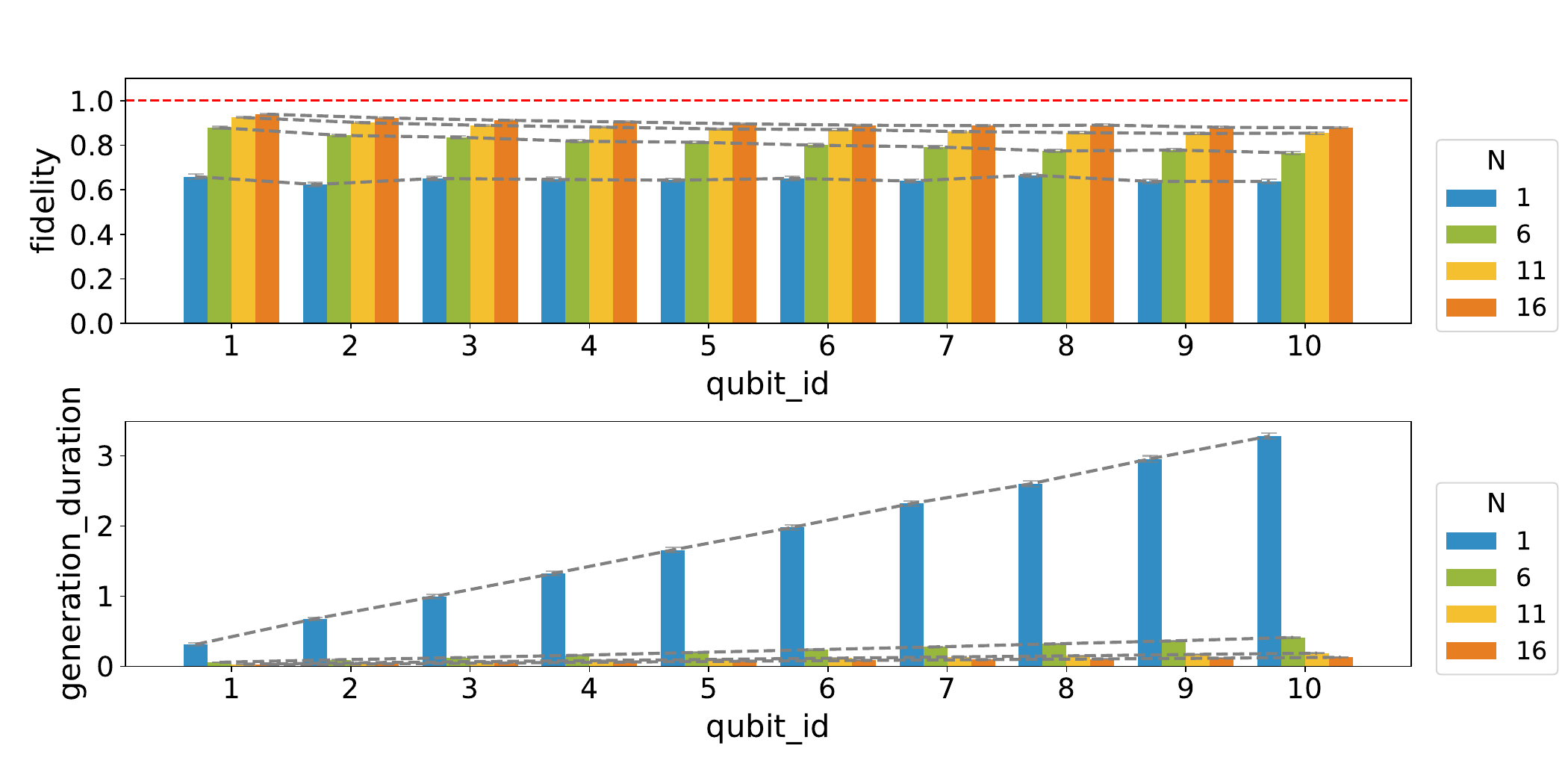}
	\caption{\label{fig:data4}Fidelity and completion time of each end-to-end entangled state provided for the request under varying network resource allocations.}
\end{figure}

Fig.~\ref{fig:data4} illustrates the fidelity and completion time of each of the ten entangled pairs during the request fulfillment process. Here, the fidelity is measured at the moment each entangled pair is successfully established, with the start time defined as the request initiation time. As shown in the figure, the earliest generated entangled pair exhibits the highest fidelity, which then gradually declines for subsequent pairs. A comparison of the data reveals that increasing the allocated resources for the request leads to a noticeable improvement in performance.

Given that the central controller has a comprehensive understanding of the performance of all nodes in the network and the utilization of overall resources, it can analyze whether the network can fulfill a given request. If feasible, the central controller can determine the optimal paths and parameters to meet the request. Specifically, the central controller can select appropriate paths based on the varying performance of nodes and the lengths of the routes.
Moreover, truncation time and resource allocation for requests are parameters that the central controller can more easily adjust. By optimizing these two parameters, it is possible to ensure that the fidelity of the final entangled state and the request completion time meet the requirements.

Currently, the simulation only models the scenario where there is a single request at any given time. Future work can involve simulating and discussing the case with multiple simultaneous requests.
\section{Summary and Outlook}\label{Outlook}
As quantum networking continues to advance, developing a comprehensive design framework for early-stage quantum networks is essential. Given the constraints of early quantum networks—such as limited resources, small-scale deployment, and suboptimal performance parameters—we propose a network architecture that supports diverse quantum applications while maintaining compatibility with three generations of quantum repeater technology. The main network is managed by a central controller, which mitigates congestion and packet loss as much as possible by adopting a connection-oriented approach, resource reservation and fixed-path routing.

Additionally, this paper introduces essential quantum network identifiers and outlines a structured workflow for processing quantum requests. We simulate the entire process of quantum requests, evaluating the impact of various parameters on the fidelity of end-to-end entangled states and request completion time. Furthermore, we examine parameters set by the central controller, including cutoff time and resource allocation strategies, and analyze their influence on overall network performance.
This work primarily establishes the foundational principles for early-stage quantum network design, without delving into specific implementation details. These details, including algorithmic strategies for central controller computations, request queuing and scheduling mechanisms, and the detailed design of quantum network identifiers and addressing schemes, will be the focus of our future research.

\bibliographystyle{ACM-Reference-Format}

\bibliography{ref1.bib}


\begin{thebibliography}{63}


\ifx \showCODEN    \undefined \def \showCODEN     #1{\unskip}     \fi
\ifx \showDOI      \undefined \def \showDOI       #1{#1}\fi
\ifx \showISBNx    \undefined \def \showISBNx     #1{\unskip}     \fi
\ifx \showISBNxiii \undefined \def \showISBNxiii  #1{\unskip}     \fi
\ifx \showISSN     \undefined \def \showISSN      #1{\unskip}     \fi
\ifx \showLCCN     \undefined \def \showLCCN      #1{\unskip}     \fi
\ifx \shownote     \undefined \def \shownote      #1{#1}          \fi
\ifx \showarticletitle \undefined \def \showarticletitle #1{#1}   \fi
\ifx \showURL      \undefined \def \showURL       {\relax}        \fi
\providecommand\bibfield[2]{#2}
\providecommand\bibinfo[2]{#2}
\providecommand\natexlab[1]{#1}
\providecommand\showeprint[2][]{arXiv:#2}

\bibitem[\protect\citeauthoryear{Avis, da~Silva, Coopmans, Dahlberg,
  Jirovsk\'a, Maier, Rabbie, Torres-Knoop, and Wehner}{Avis
  et~al\mbox{.}}{2023a}]%
        {Avis_requirements_2023}
\bibfield{author}{\bibinfo{person}{Guus Avis},
  \bibinfo{person}{Francisco~Ferreira da Silva}, \bibinfo{person}{Tim
  Coopmans}, \bibinfo{person}{Axel Dahlberg}, \bibinfo{person}{Hana
  Jirovsk\'a}, \bibinfo{person}{David Maier}, \bibinfo{person}{Julian Rabbie},
  \bibinfo{person}{Ariana Torres-Knoop}, {and} \bibinfo{person}{Stephanie
  Wehner}.} \bibinfo{year}{2023}\natexlab{a}.
\newblock \showarticletitle{{Requirements for a processing-node quantum
  repeater on a real-world fiber grid}}.
\newblock \bibinfo{journal}{{\em npj Quantum Inf.\/}} \bibinfo{volume}{9},
  \bibinfo{number}{1} (\bibinfo{year}{2023}), \bibinfo{pages}{100}.
\newblock
\showDOI{%
\url{https://doi.org/10.1038/s41534-023-00765-x}}
\showeprint[arxiv]{quant-ph/2207.10579}


\bibitem[\protect\citeauthoryear{Avis, Rozp{\k{e}}{\k{e}}dek, and Wehner}{Avis
  et~al\mbox{.}}{2023b}]%
        {avis_analysis_2023}
\bibfield{author}{\bibinfo{person}{Guus Avis}, \bibinfo{person}{Filip
  Rozp{\k{e}}{\k{e}}dek}, {and} \bibinfo{person}{Stephanie Wehner}.}
  \bibinfo{year}{2023}\natexlab{b}.
\newblock \showarticletitle{Analysis of Multipartite Entanglement Distribution
  Using a Central Quantum-Network Node}.
\newblock \bibinfo{journal}{{\em Phys. Rev. A\/}} \bibinfo{volume}{107},
  \bibinfo{number}{1} (\bibinfo{date}{Jan.} \bibinfo{year}{2023}),
  \bibinfo{pages}{012609}.
\newblock
\showDOI{%
\url{https://doi.org/10.1103/PhysRevA.107.012609}}


\bibitem[\protect\citeauthoryear{Azuma, B{\"a}uml, Coopmans, Elkouss, and
  Li}{Azuma et~al\mbox{.}}{2021}]%
        {azuma_tools_2021}
\bibfield{author}{\bibinfo{person}{Koji Azuma}, \bibinfo{person}{Stefan
  B{\"a}uml}, \bibinfo{person}{Tim Coopmans}, \bibinfo{person}{David Elkouss},
  {and} \bibinfo{person}{Boxi Li}.} \bibinfo{year}{2021}\natexlab{}.
\newblock \showarticletitle{Tools for Quantum Network Design}.
\newblock \bibinfo{journal}{{\em AVS Quantum Sci.\/}} \bibinfo{volume}{3},
  \bibinfo{number}{1} (\bibinfo{date}{March} \bibinfo{year}{2021}),
  \bibinfo{pages}{014101}.
\newblock
\showISSN{2639-0213}
\showDOI{%
\url{https://doi.org/10.1116/5.0024062}}


\bibitem[\protect\citeauthoryear{Beals, Brierley, Gray, Harrow, Kutin, Linden,
  Shepherd, and Stather}{Beals et~al\mbox{.}}{2013}]%
        {beals_efficient_2013}
\bibfield{author}{\bibinfo{person}{Robert Beals}, \bibinfo{person}{Stephen
  Brierley}, \bibinfo{person}{Oliver Gray}, \bibinfo{person}{Aram~W. Harrow},
  \bibinfo{person}{Samuel Kutin}, \bibinfo{person}{Noah Linden},
  \bibinfo{person}{Dan Shepherd}, {and} \bibinfo{person}{Mark Stather}.}
  \bibinfo{year}{2013}\natexlab{}.
\newblock \showarticletitle{Efficient Distributed Quantum Computing}.
\newblock \bibinfo{journal}{{\em Proc. R. Soc. A.\/}} \bibinfo{volume}{469},
  \bibinfo{number}{2153} (\bibinfo{date}{May} \bibinfo{year}{2013}),
  \bibinfo{pages}{20120686}.
\newblock
\showISSN{1364-5021, 1471-2946}
\showDOI{%
\url{https://doi.org/10.1098/rspa.2012.0686}}


\bibitem[\protect\citeauthoryear{Bengtsson and {\.Z}yczkowski}{Bengtsson and
  {\.Z}yczkowski}{2017}]%
        {bengtsson_geometry_2017}
\bibfield{author}{\bibinfo{person}{Ingemar Bengtsson} {and}
  \bibinfo{person}{Karol {\.Z}yczkowski}.} \bibinfo{year}{2017}\natexlab{}.
\newblock \bibinfo{booktitle}{{\em Geometry of Quantum States: An Introduction
  to Quantum Entanglement}}.
\newblock \bibinfo{publisher}{Cambridge university press}.
\newblock


\bibitem[\protect\citeauthoryear{Bennett, Brassard, Cr{\'e}peau, Jozsa, Peres,
  and Wootters}{Bennett et~al\mbox{.}}{1993}]%
        {bennett_teleporting_1993}
\bibfield{author}{\bibinfo{person}{Charles~H. Bennett}, \bibinfo{person}{Gilles
  Brassard}, \bibinfo{person}{Claude Cr{\'e}peau}, \bibinfo{person}{Richard
  Jozsa}, \bibinfo{person}{Asher Peres}, {and} \bibinfo{person}{William~K.
  Wootters}.} \bibinfo{year}{1993}\natexlab{}.
\newblock \showarticletitle{Teleporting an Unknown Quantum State via Dual
  Classical and Einstein-Podolsky-Rosen Channels}.
\newblock \bibinfo{journal}{{\em Phys. Rev. Lett.\/}} \bibinfo{volume}{70},
  \bibinfo{number}{13} (\bibinfo{date}{March} \bibinfo{year}{1993}),
  \bibinfo{pages}{1895--1899}.
\newblock
\showISSN{0031-9007}
\showDOI{%
\url{https://doi.org/10.1103/PhysRevLett.70.1895}}


\bibitem[\protect\citeauthoryear{Bennett and DiVincenzo}{Bennett and
  DiVincenzo}{2000}]%
        {bennett_quantum_2000}
\bibfield{author}{\bibinfo{person}{Charles~H. Bennett} {and}
  \bibinfo{person}{David~P. DiVincenzo}.} \bibinfo{year}{2000}\natexlab{}.
\newblock \showarticletitle{Quantum Information and Computation}.
\newblock \bibinfo{journal}{{\em Nature\/}} \bibinfo{volume}{404},
  \bibinfo{number}{6775} (\bibinfo{date}{March} \bibinfo{year}{2000}),
  \bibinfo{pages}{247--255}.
\newblock
\showISSN{1476-4687}
\showDOI{%
\url{https://doi.org/10.1038/35005001}}


\bibitem[\protect\citeauthoryear{Brunner, Cavalcanti, Pironio, Scarani, and
  Wehner}{Brunner et~al\mbox{.}}{2014}]%
        {brunner_bell_2014}
\bibfield{author}{\bibinfo{person}{Nicolas Brunner}, \bibinfo{person}{Daniel
  Cavalcanti}, \bibinfo{person}{Stefano Pironio}, \bibinfo{person}{Valerio
  Scarani}, {and} \bibinfo{person}{Stephanie Wehner}.}
  \bibinfo{year}{2014}\natexlab{}.
\newblock \showarticletitle{Bell Nonlocality}.
\newblock \bibinfo{journal}{{\em Rev. Mod. Phys.\/}} \bibinfo{volume}{86},
  \bibinfo{number}{2} (\bibinfo{date}{April} \bibinfo{year}{2014}),
  \bibinfo{pages}{419--478}.
\newblock
\showISSN{0034-6861, 1539-0756}
\showDOI{%
\url{https://doi.org/10.1103/RevModPhys.86.419}}


\bibitem[\protect\citeauthoryear{Buhrman and R{\"o}hrig}{Buhrman and
  R{\"o}hrig}{2003}]%
        {buhrman_distributed_2003}
\bibfield{author}{\bibinfo{person}{Harry Buhrman} {and} \bibinfo{person}{Hein
  R{\"o}hrig}.} \bibinfo{year}{2003}\natexlab{}.
\newblock \showarticletitle{Distributed Quantum Computing}. In
  \bibinfo{booktitle}{{\em Mathematical Foundations of Computer Science 2003}},
  \bibfield{editor}{\bibinfo{person}{Branislav Rovan} {and}
  \bibinfo{person}{Peter Vojt{\'a}{\v s}}} (Eds.). \bibinfo{publisher}{Springer
  Berlin Heidelberg}, \bibinfo{address}{Berlin, Heidelberg},
  \bibinfo{pages}{1--20}.
\newblock
\showISBNx{978-3-540-45138-9}


\bibitem[\protect\citeauthoryear{Chehimi and Saad}{Chehimi and Saad}{2022}]%
        {chehimi_physics-informed_2022}
\bibfield{author}{\bibinfo{person}{Mahdi Chehimi} {and} \bibinfo{person}{Walid
  Saad}.} \bibinfo{year}{2022}\natexlab{}.
\newblock \showarticletitle{Physics-Informed Quantum Communication Networks: A
  Vision Toward the Quantum Internet}.
\newblock \bibinfo{journal}{{\em IEEE Netw.\/}} \bibinfo{volume}{36},
  \bibinfo{number}{5} (\bibinfo{date}{Sept.} \bibinfo{year}{2022}),
  \bibinfo{pages}{32--38}.
\newblock
\showISSN{0890-8044, 1558-156X}
\showDOI{%
\url{https://doi.org/10.1109/MNET.001.2200153}}


\bibitem[\protect\citeauthoryear{Chen, Zhang, Chen, Cai, Liao, Zhang, Chen,
  Yin, Ren, Chen, Han, Yu, Liang, Zhou, Yuan, Zhao, Wang, Jiang, Zhang, Liu,
  Li, Shen, Cao, Lu, Shu, Wang, Li, Liu, Xu, Wang, Peng, and Pan}{Chen
  et~al\mbox{.}}{2021}]%
        {chen_integrated_2021}
\bibfield{author}{\bibinfo{person}{Yu-Ao Chen}, \bibinfo{person}{Qiang Zhang},
  \bibinfo{person}{Teng-Yun Chen}, \bibinfo{person}{Wen-Qi Cai},
  \bibinfo{person}{Sheng-Kai Liao}, \bibinfo{person}{Jun Zhang},
  \bibinfo{person}{Kai Chen}, \bibinfo{person}{Juan Yin},
  \bibinfo{person}{Ji-Gang Ren}, \bibinfo{person}{Zhu Chen},
  \bibinfo{person}{Sheng-Long Han}, \bibinfo{person}{Qing Yu},
  \bibinfo{person}{Ken Liang}, \bibinfo{person}{Fei Zhou},
  \bibinfo{person}{Xiao Yuan}, \bibinfo{person}{Mei-Sheng Zhao},
  \bibinfo{person}{Tian-Yin Wang}, \bibinfo{person}{Xiao Jiang},
  \bibinfo{person}{Liang Zhang}, \bibinfo{person}{Wei-Yue Liu},
  \bibinfo{person}{Yang Li}, \bibinfo{person}{Qi Shen}, \bibinfo{person}{Yuan
  Cao}, \bibinfo{person}{Chao-Yang Lu}, \bibinfo{person}{Rong Shu},
  \bibinfo{person}{Jian-Yu Wang}, \bibinfo{person}{Li Li},
  \bibinfo{person}{Nai-Le Liu}, \bibinfo{person}{Feihu Xu},
  \bibinfo{person}{Xiang-Bin Wang}, \bibinfo{person}{Cheng-Zhi Peng}, {and}
  \bibinfo{person}{Jian-Wei Pan}.} \bibinfo{year}{2021}\natexlab{}.
\newblock \showarticletitle{An Integrated Space-to-Ground Quantum Communication
  Network over 4,600 Kilometres}.
\newblock \bibinfo{journal}{{\em Nature\/}} \bibinfo{volume}{589},
  \bibinfo{number}{7841} (\bibinfo{date}{Jan.} \bibinfo{year}{2021}),
  \bibinfo{pages}{214--219}.
\newblock
\showISSN{0028-0836, 1476-4687}
\showDOI{%
\url{https://doi.org/10.1038/s41586-020-03093-8}}


\bibitem[\protect\citeauthoryear{Cuomo, Caleffi, and Cacciapuoti}{Cuomo
  et~al\mbox{.}}{2020}]%
        {cuomo_towards_2020}
\bibfield{author}{\bibinfo{person}{Daniele Cuomo}, \bibinfo{person}{Marcello
  Caleffi}, {and} \bibinfo{person}{Angela~Sara Cacciapuoti}.}
  \bibinfo{year}{2020}\natexlab{}.
\newblock \showarticletitle{Towards a Distributed Quantum Computing Ecosystem}.
\newblock \bibinfo{journal}{{\em IET Quantum Commun.\/}} \bibinfo{volume}{1},
  \bibinfo{number}{1} (\bibinfo{date}{July} \bibinfo{year}{2020}),
  \bibinfo{pages}{3--8}.
\newblock
\showISSN{2632-8925, 2632-8925}
\showDOI{%
\url{https://doi.org/10.1049/iet-qtc.2020.0002}}


\bibitem[\protect\citeauthoryear{Dahlberg, Skrzypczyk, Coopmans, Wubben,
  Rozpundefineddek, Pompili, Stolk, Pawe{\l}czak, Knegjens, {de Oliveira
  Filho}, Hanson, and Wehner}{Dahlberg et~al\mbox{.}}{2019}]%
        {dahlberg_link_2019}
\bibfield{author}{\bibinfo{person}{Axel Dahlberg}, \bibinfo{person}{Matthew
  Skrzypczyk}, \bibinfo{person}{Tim Coopmans}, \bibinfo{person}{Leon Wubben},
  \bibinfo{person}{Filip Rozpundefineddek}, \bibinfo{person}{Matteo Pompili},
  \bibinfo{person}{Arian Stolk}, \bibinfo{person}{Przemys{\l}aw Pawe{\l}czak},
  \bibinfo{person}{Robert Knegjens}, \bibinfo{person}{Julio {de Oliveira
  Filho}}, \bibinfo{person}{Ronald Hanson}, {and} \bibinfo{person}{Stephanie
  Wehner}.} \bibinfo{year}{2019}\natexlab{}.
\newblock \showarticletitle{A Link Layer Protocol for Quantum Networks}. In
  \bibinfo{booktitle}{{\em Proceedings of the ACM Special Interest Group on
  Data Communication}} {\em (\bibinfo{series}{SIGCOMM '19})}.
  \bibinfo{publisher}{Association for Computing Machinery},
  \bibinfo{address}{New York, NY, USA}, \bibinfo{pages}{159--173}.
\newblock
\showISBNx{978-1-4503-5956-6}
\showDOI{%
\url{https://doi.org/10.1145/3341302.3342070}}


\bibitem[\protect\citeauthoryear{Degen, Reinhard, and Cappellaro}{Degen
  et~al\mbox{.}}{2017}]%
        {degen_quantum_2017}
\bibfield{author}{\bibinfo{person}{C.~L. Degen}, \bibinfo{person}{F. Reinhard},
  {and} \bibinfo{person}{P. Cappellaro}.} \bibinfo{year}{2017}\natexlab{}.
\newblock \showarticletitle{Quantum Sensing}.
\newblock \bibinfo{journal}{{\em Rev. Mod. Phys.\/}} \bibinfo{volume}{89},
  \bibinfo{number}{3} (\bibinfo{date}{July} \bibinfo{year}{2017}),
  \bibinfo{pages}{035002}.
\newblock
\showISSN{0034-6861, 1539-0756}
\showDOI{%
\url{https://doi.org/10.1103/RevModPhys.89.035002}}


\bibitem[\protect\citeauthoryear{Devitt, Nemoto, and Munro}{Devitt
  et~al\mbox{.}}{2013}]%
        {devitt_quantum_2013}
\bibfield{author}{\bibinfo{person}{Simon~J. Devitt}, \bibinfo{person}{Kae
  Nemoto}, {and} \bibinfo{person}{William~J. Munro}.}
  \bibinfo{year}{2013}\natexlab{}.
\newblock \showarticletitle{Quantum Error Correction for Beginners}.
\newblock \bibinfo{journal}{{\em Rep. Prog. Phys.\/}} \bibinfo{volume}{76},
  \bibinfo{number}{7} (\bibinfo{date}{July} \bibinfo{year}{2013}),
  \bibinfo{pages}{076001}.
\newblock
\showISSN{0034-4885, 1361-6633}
\showDOI{%
\url{https://doi.org/10.1088/0034-4885/76/7/076001}}


\bibitem[\protect\citeauthoryear{D{\"u}r, Lamprecht, and Heusler}{D{\"u}r
  et~al\mbox{.}}{2017}]%
        {Dur_towards_2017}
\bibfield{author}{\bibinfo{person}{Wolfgang D{\"u}r}, \bibinfo{person}{Raphael
  Lamprecht}, {and} \bibinfo{person}{Stefan Heusler}.}
  \bibinfo{year}{2017}\natexlab{}.
\newblock \showarticletitle{Towards a Quantum Internet}.
\newblock \bibinfo{journal}{{\em Eur. J. Phys.\/}} \bibinfo{volume}{38},
  \bibinfo{number}{4} (\bibinfo{date}{May} \bibinfo{year}{2017}),
  \bibinfo{pages}{043001}.
\newblock
\showDOI{%
\url{https://doi.org/10.1088/1361-6404/aa6df7}}


\bibitem[\protect\citeauthoryear{Dynes, Wonfor, Tam, Sharpe, Takahashi,
  Lucamarini, Plews, Yuan, Dixon, Cho, Tanizawa, Elbers, Grei{\ss}er, White,
  Penty, and Shields}{Dynes et~al\mbox{.}}{2019}]%
        {dynes_cambridge_2019}
\bibfield{author}{\bibinfo{person}{J.~F. Dynes}, \bibinfo{person}{A. Wonfor},
  \bibinfo{person}{W.~W.~S. Tam}, \bibinfo{person}{A.~W. Sharpe},
  \bibinfo{person}{R. Takahashi}, \bibinfo{person}{M. Lucamarini},
  \bibinfo{person}{A. Plews}, \bibinfo{person}{Z.~L. Yuan},
  \bibinfo{person}{A.~R. Dixon}, \bibinfo{person}{J. Cho}, \bibinfo{person}{Y.
  Tanizawa}, \bibinfo{person}{J.~P. Elbers}, \bibinfo{person}{H. Grei{\ss}er},
  \bibinfo{person}{I.~H. White}, \bibinfo{person}{R.~V. Penty}, {and}
  \bibinfo{person}{A.~J. Shields}.} \bibinfo{year}{2019}\natexlab{}.
\newblock \showarticletitle{Cambridge Quantum Network}.
\newblock \bibinfo{journal}{{\em npj Quantum Inf.\/}} \bibinfo{volume}{5},
  \bibinfo{number}{1} (\bibinfo{date}{Nov.} \bibinfo{year}{2019}),
  \bibinfo{pages}{101}.
\newblock
\showISSN{2056-6387}
\showDOI{%
\url{https://doi.org/10.1038/s41534-019-0221-4}}


\bibitem[\protect\citeauthoryear{Eldredge, {Foss-Feig}, Gross, Rolston, and
  Gorshkov}{Eldredge et~al\mbox{.}}{2018}]%
        {eldredge_optimal_2018}
\bibfield{author}{\bibinfo{person}{Zachary Eldredge}, \bibinfo{person}{Michael
  {Foss-Feig}}, \bibinfo{person}{Jonathan~A. Gross}, \bibinfo{person}{S.~L.
  Rolston}, {and} \bibinfo{person}{Alexey~V. Gorshkov}.}
  \bibinfo{year}{2018}\natexlab{}.
\newblock \showarticletitle{Optimal and Secure Measurement Protocols for
  Quantum Sensor Networks}.
\newblock \bibinfo{journal}{{\em Phys. Rev. A\/}} \bibinfo{volume}{97},
  \bibinfo{number}{4} (\bibinfo{date}{April} \bibinfo{year}{2018}),
  \bibinfo{pages}{042337}.
\newblock
\showISSN{2469-9926, 2469-9934}
\showDOI{%
\url{https://doi.org/10.1103/PhysRevA.97.042337}}


\bibitem[\protect\citeauthoryear{Fang, Zhao, Li, Li, and Duan}{Fang
  et~al\mbox{.}}{2023}]%
        {fang_quantum_2022}
\bibfield{author}{\bibinfo{person}{Kun Fang}, \bibinfo{person}{Jingtian Zhao},
  \bibinfo{person}{Xiufan Li}, \bibinfo{person}{Yifei Li}, {and}
  \bibinfo{person}{Runyao Duan}.} \bibinfo{year}{2023}\natexlab{}.
\newblock \showarticletitle{Quantum NETwork: From Theory to Practice}.
\newblock \bibinfo{journal}{{\em Science China Information Sciences\/}}
  \bibinfo{volume}{66}, \bibinfo{number}{8} (\bibinfo{date}{July}
  \bibinfo{year}{2023}), \bibinfo{pages}{180509}.
\newblock
\showISSN{1869-1919}
\showDOI{%
\url{https://doi.org/10.1007/s11432-023-3773-4}}


\bibitem[\protect\citeauthoryear{Gisin and Thew}{Gisin and Thew}{2007}]%
        {gisin_quantum_2007}
\bibfield{author}{\bibinfo{person}{Nicolas Gisin} {and} \bibinfo{person}{Rob
  Thew}.} \bibinfo{year}{2007}\natexlab{}.
\newblock \showarticletitle{Quantum Communication}.
\newblock \bibinfo{journal}{{\em Nat. Photon.\/}} \bibinfo{volume}{1},
  \bibinfo{number}{3} (\bibinfo{date}{March} \bibinfo{year}{2007}),
  \bibinfo{pages}{165--171}.
\newblock
\showISSN{1749-4893}
\showDOI{%
\url{https://doi.org/10.1038/nphoton.2007.22}}


\bibitem[\protect\citeauthoryear{Guo, Breum, Borregaard, Izumi, Larsen,
  Gehring, Christandl, {Neergaard-Nielsen}, and Andersen}{Guo
  et~al\mbox{.}}{2020}]%
        {guo_distributed_2020}
\bibfield{author}{\bibinfo{person}{Xueshi Guo}, \bibinfo{person}{Casper~R.
  Breum}, \bibinfo{person}{Johannes Borregaard}, \bibinfo{person}{Shuro Izumi},
  \bibinfo{person}{Mikkel~V. Larsen}, \bibinfo{person}{Tobias Gehring},
  \bibinfo{person}{Matthias Christandl}, \bibinfo{person}{Jonas~S.
  {Neergaard-Nielsen}}, {and} \bibinfo{person}{Ulrik~L. Andersen}.}
  \bibinfo{year}{2020}\natexlab{}.
\newblock \showarticletitle{Distributed Quantum Sensing in a
  Continuous-Variable Entangled Network}.
\newblock \bibinfo{journal}{{\em Nat. Phys.\/}} \bibinfo{volume}{16},
  \bibinfo{number}{3} (\bibinfo{date}{March} \bibinfo{year}{2020}),
  \bibinfo{pages}{281--284}.
\newblock
\showISSN{1745-2473, 1745-2481}
\showDOI{%
\url{https://doi.org/10.1038/s41567-019-0743-x}}


\bibitem[\protect\citeauthoryear{Gyongyosi and Imre}{Gyongyosi and
  Imre}{2019}]%
        {gyongyosi_survey_2019}
\bibfield{author}{\bibinfo{person}{Laszlo Gyongyosi} {and}
  \bibinfo{person}{Sandor Imre}.} \bibinfo{year}{2019}\natexlab{}.
\newblock \showarticletitle{A Survey on Quantum Computing Technology}.
\newblock \bibinfo{journal}{{\em Comput. Sci. Rev.\/}}  \bibinfo{volume}{31}
  (\bibinfo{date}{Feb.} \bibinfo{year}{2019}), \bibinfo{pages}{51--71}.
\newblock
\showISSN{15740137}
\showDOI{%
\url{https://doi.org/10.1016/j.cosrev.2018.11.002}}


\bibitem[\protect\citeauthoryear{Hahn, Pappa, and Eisert}{Hahn
  et~al\mbox{.}}{2019}]%
        {hahn_quantum_2019}
\bibfield{author}{\bibinfo{person}{F. Hahn}, \bibinfo{person}{A. Pappa}, {and}
  \bibinfo{person}{J. Eisert}.} \bibinfo{year}{2019}\natexlab{}.
\newblock \showarticletitle{Quantum Network Routing and Local Complementation}.
\newblock \bibinfo{journal}{{\em npj Quantum Inf.\/}} \bibinfo{volume}{5},
  \bibinfo{number}{1} (\bibinfo{date}{Dec.} \bibinfo{year}{2019}),
  \bibinfo{pages}{76}.
\newblock
\showISSN{2056-6387}
\showDOI{%
\url{https://doi.org/10.1038/s41534-019-0191-6}}


\bibitem[\protect\citeauthoryear{He, Zhang, Loke, Lin, and Lu}{He
  et~al\mbox{.}}{2024}]%
        {he_building_2024}
\bibfield{author}{\bibinfo{person}{Binjie He}, \bibinfo{person}{Dong Zhang},
  \bibinfo{person}{Seng~W. Loke}, \bibinfo{person}{Shengrui Lin}, {and}
  \bibinfo{person}{Luke Lu}.} \bibinfo{year}{2024}\natexlab{}.
\newblock \showarticletitle{Building a Hierarchical Architecture and
  Communication Model for the Quantum Internet}.
\newblock \bibinfo{journal}{{\em IEEE Journal on Selected Areas in
  Communications\/}} \bibinfo{volume}{42}, \bibinfo{number}{7}
  (\bibinfo{year}{2024}), \bibinfo{pages}{1919--1935}.
\newblock
\showDOI{%
\url{https://doi.org/10.1109/JSAC.2024.3380103}}


\bibitem[\protect\citeauthoryear{Horodecki, Horodecki, Horodecki, and
  Horodecki}{Horodecki et~al\mbox{.}}{2009}]%
        {horodecki_quantum_2009}
\bibfield{author}{\bibinfo{person}{Ryszard Horodecki},
  \bibinfo{person}{Pawe{\l} Horodecki}, \bibinfo{person}{Micha{\l} Horodecki},
  {and} \bibinfo{person}{Karol Horodecki}.} \bibinfo{year}{2009}\natexlab{}.
\newblock \showarticletitle{Quantum Entanglement}.
\newblock \bibinfo{journal}{{\em Rev. Mod. Phys.\/}} \bibinfo{volume}{81},
  \bibinfo{number}{2} (\bibinfo{date}{June} \bibinfo{year}{2009}),
  \bibinfo{pages}{865--942}.
\newblock
\showISSN{0034-6861, 1539-0756}
\showDOI{%
\url{https://doi.org/10.1103/RevModPhys.81.865}}


\bibitem[\protect\citeauthoryear{Illiano, Caleffi, Manzalini, and
  Cacciapuoti}{Illiano et~al\mbox{.}}{2022}]%
        {illiano_quantum_2022}
\bibfield{author}{\bibinfo{person}{Jessica Illiano}, \bibinfo{person}{Marcello
  Caleffi}, \bibinfo{person}{Antonio Manzalini}, {and}
  \bibinfo{person}{Angela~Sara Cacciapuoti}.} \bibinfo{year}{2022}\natexlab{}.
\newblock \showarticletitle{Quantum Internet Protocol Stack: A Comprehensive
  Survey}.
\newblock \bibinfo{journal}{{\em Comput. Netw.\/}}  \bibinfo{volume}{213}
  (\bibinfo{year}{2022}), \bibinfo{pages}{109092}.
\newblock
\showISSN{1389-1286}
\showDOI{%
\url{https://doi.org/10.1016/j.comnet.2022.109092}}


\bibitem[\protect\citeauthoryear{Kimble}{Kimble}{2008}]%
        {kimble_quantum_2008}
\bibfield{author}{\bibinfo{person}{H.~J. Kimble}.}
  \bibinfo{year}{2008}\natexlab{}.
\newblock \showarticletitle{The Quantum Internet}.
\newblock \bibinfo{journal}{{\em Nature\/}} \bibinfo{volume}{453},
  \bibinfo{number}{7198} (\bibinfo{date}{June} \bibinfo{year}{2008}),
  \bibinfo{pages}{1023--1030}.
\newblock
\showISSN{0028-0836, 1476-4687}
\showDOI{%
\url{https://doi.org/10.1038/nature07127}}


\bibitem[\protect\citeauthoryear{Kozlowski, Dahlberg, and Wehner}{Kozlowski
  et~al\mbox{.}}{2020}]%
        {kozlowski_designing_2020}
\bibfield{author}{\bibinfo{person}{Wojciech Kozlowski}, \bibinfo{person}{Axel
  Dahlberg}, {and} \bibinfo{person}{Stephanie Wehner}.}
  \bibinfo{year}{2020}\natexlab{}.
\newblock \showarticletitle{Designing a Quantum Network Protocol}. In
  \bibinfo{booktitle}{{\em Proceedings of the 16th International Conference on
  Emerging Networking EXperiments and Technologies}}. \bibinfo{publisher}{ACM},
  \bibinfo{address}{Barcelona Spain}, \bibinfo{pages}{1--16}.
\newblock
\showISBNx{978-1-4503-7948-9}
\showDOI{%
\url{https://doi.org/10.1145/3386367.3431293}}


\bibitem[\protect\citeauthoryear{Kozlowski and Wehner}{Kozlowski and
  Wehner}{2019}]%
        {kozlowski_towards_2019}
\bibfield{author}{\bibinfo{person}{Wojciech Kozlowski} {and}
  \bibinfo{person}{Stephanie Wehner}.} \bibinfo{year}{2019}\natexlab{}.
\newblock \showarticletitle{Towards Large-Scale Quantum Networks}. In
  \bibinfo{booktitle}{{\em Proceedings of the Sixth Annual ACM International
  Conference on Nanoscale Computing and Communication}}. \bibinfo{pages}{1--7}.
\newblock
\showDOI{%
\url{https://doi.org/10.1145/3345312.3345497}}


\bibitem[\protect\citeauthoryear{Ladd, Jelezko, Laflamme, Nakamura, Monroe, and
  O'Brien}{Ladd et~al\mbox{.}}{2010}]%
        {ladd_quantum_2010}
\bibfield{author}{\bibinfo{person}{T.~D. Ladd}, \bibinfo{person}{F. Jelezko},
  \bibinfo{person}{R. Laflamme}, \bibinfo{person}{Y. Nakamura},
  \bibinfo{person}{C. Monroe}, {and} \bibinfo{person}{J.~L. O'Brien}.}
  \bibinfo{year}{2010}\natexlab{}.
\newblock \showarticletitle{Quantum Computers}.
\newblock \bibinfo{journal}{{\em Nature\/}} \bibinfo{volume}{464},
  \bibinfo{number}{7285} (\bibinfo{date}{March} \bibinfo{year}{2010}),
  \bibinfo{pages}{45--53}.
\newblock
\showISSN{1476-4687}
\showDOI{%
\url{https://doi.org/10.1038/nature08812}}


\bibitem[\protect\citeauthoryear{Li, Li, Liu, and Cappellaro}{Li
  et~al\mbox{.}}{2021a}]%
        {li_effective_2021}
\bibfield{author}{\bibinfo{person}{Changhao Li}, \bibinfo{person}{Tianyi Li},
  \bibinfo{person}{Yi-Xiang Liu}, {and} \bibinfo{person}{Paola Cappellaro}.}
  \bibinfo{year}{2021}\natexlab{a}.
\newblock \showarticletitle{Effective Routing Design for Remote Entanglement
  Generation on Quantum Networks}.
\newblock \bibinfo{journal}{{\em npj Quantum Inf.\/}} \bibinfo{volume}{7},
  \bibinfo{number}{1} (\bibinfo{date}{Dec.} \bibinfo{year}{2021}),
  \bibinfo{pages}{10}.
\newblock
\showISSN{2056-6387}
\showDOI{%
\url{https://doi.org/10.1038/s41534-020-00344-4}}


\bibitem[\protect\citeauthoryear{Li, Wang, Xue, Li, Yu, Sun, and Lu}{Li
  et~al\mbox{.}}{2022}]%
        {li_fidelity-guaranteed_2022}
\bibfield{author}{\bibinfo{person}{Jian Li}, \bibinfo{person}{Mingjun Wang},
  \bibinfo{person}{Kaiping Xue}, \bibinfo{person}{Ruidong Li},
  \bibinfo{person}{Nenghai Yu}, \bibinfo{person}{Qibin Sun}, {and}
  \bibinfo{person}{Jun Lu}.} \bibinfo{year}{2022}\natexlab{}.
\newblock \showarticletitle{Fidelity-Guaranteed Entanglement Routing in Quantum
  Networks}.
\newblock \bibinfo{journal}{{\em IEEE Trans. Commun.\/}}
  (\bibinfo{year}{2022}), \bibinfo{pages}{1--1}.
\newblock
\showDOI{%
\url{https://doi.org/10.1109/TCOMM.2022.3200115}}


\bibitem[\protect\citeauthoryear{Li, Zhang, Zhang, Huang, and Yu}{Li
  et~al\mbox{.}}{2024}]%
        {li_survey_2024}
\bibfield{author}{\bibinfo{person}{Yuan Li}, \bibinfo{person}{Hao Zhang},
  \bibinfo{person}{Chen Zhang}, \bibinfo{person}{Tao Huang}, {and}
  \bibinfo{person}{F.~Richard Yu}.} \bibinfo{year}{2024}\natexlab{}.
\newblock \showarticletitle{A Survey of Quantum Internet Protocols From a
  Layered Perspective}.
\newblock \bibinfo{journal}{{\em IEEE Communications Surveys \& Tutorials\/}}
  \bibinfo{volume}{26}, \bibinfo{number}{3} (\bibinfo{year}{2024}),
  \bibinfo{pages}{1606--1634}.
\newblock
\showDOI{%
\url{https://doi.org/10.1109/COMST.2024.3361662}}


\bibitem[\protect\citeauthoryear{Li, Xue, Li, Chen, Li, Wang, Yu, Wei, Sun, and
  Lu}{Li et~al\mbox{.}}{2023}]%
        {li_entanglement_2023}
\bibfield{author}{\bibinfo{person}{Zhonghui Li}, \bibinfo{person}{Kaiping Xue},
  \bibinfo{person}{Jian Li}, \bibinfo{person}{Lutong Chen},
  \bibinfo{person}{Ruidong Li}, \bibinfo{person}{Zhaoying Wang},
  \bibinfo{person}{Nenghai Yu}, \bibinfo{person}{David~S.L. Wei},
  \bibinfo{person}{Qibin Sun}, {and} \bibinfo{person}{Jun Lu}.}
  \bibinfo{year}{2023}\natexlab{}.
\newblock \showarticletitle{Entanglement-Assisted Quantum Networks: Mechanics,
  Enabling Technologies, Challenges, and Research Directions}.
\newblock \bibinfo{journal}{{\em IEEE Commun. Surveys Tuts.\/}}
  (\bibinfo{year}{2023}), \bibinfo{pages}{1--1}.
\newblock
\showDOI{%
\url{https://doi.org/10.1109/COMST.2023.3294240}}


\bibitem[\protect\citeauthoryear{Li, Xue, Li, Yu, Liu, Wei, Sun, and Lu}{Li
  et~al\mbox{.}}{2021b}]%
        {li_building_2021}
\bibfield{author}{\bibinfo{person}{Zhonghui Li}, \bibinfo{person}{Kaiping Xue},
  \bibinfo{person}{Jian Li}, \bibinfo{person}{Nenghai Yu},
  \bibinfo{person}{Jianqing Liu}, \bibinfo{person}{David S.~L. Wei},
  \bibinfo{person}{Qibin Sun}, {and} \bibinfo{person}{Jun Lu}.}
  \bibinfo{year}{2021}\natexlab{b}.
\newblock \showarticletitle{Building a Large-Scale and Wide-Area Quantum
  Internet Based on an OSI-Alike Model}.
\newblock \bibinfo{journal}{{\em China Commun.\/}} \bibinfo{volume}{18},
  \bibinfo{number}{10} (\bibinfo{year}{2021}), \bibinfo{pages}{1--14}.
\newblock
\showDOI{%
\url{https://doi.org/10.23919/JCC.2021.10.001}}


\bibitem[\protect\citeauthoryear{Liu, Allcock, Cai, Zhang, and Lui}{Liu
  et~al\mbox{.}}{2022}]%
        {liu_quantum_2022}
\bibfield{author}{\bibinfo{person}{Maoli Liu}, \bibinfo{person}{Jonathan
  Allcock}, \bibinfo{person}{Kechao Cai}, \bibinfo{person}{Shengyu Zhang},
  {and} \bibinfo{person}{John~C.S. Lui}.} \bibinfo{year}{2022}\natexlab{}.
\newblock \showarticletitle{Quantum Networks with Multiple Service Providers:
  Transport Layer Protocols and Research Opportunities}.
\newblock \bibinfo{journal}{{\em IEEE Netw.\/}} \bibinfo{volume}{36},
  \bibinfo{number}{5} (\bibinfo{date}{Sept.} \bibinfo{year}{2022}),
  \bibinfo{pages}{56--62}.
\newblock
\showISSN{0890-8044, 1558-156X}
\showDOI{%
\url{https://doi.org/10.1109/MNET.001.2200151}}


\bibitem[\protect\citeauthoryear{Malia, Wu, {Mart{\'i}nez-Rinc{\'o}n}, and
  Kasevich}{Malia et~al\mbox{.}}{2022}]%
        {malia_distributed_2022}
\bibfield{author}{\bibinfo{person}{Benjamin~K. Malia}, \bibinfo{person}{Yunfan
  Wu}, \bibinfo{person}{Juli{\'a}n {Mart{\'i}nez-Rinc{\'o}n}}, {and}
  \bibinfo{person}{Mark~A. Kasevich}.} \bibinfo{year}{2022}\natexlab{}.
\newblock \showarticletitle{Distributed Quantum Sensing with Mode-Entangled
  Spin-Squeezed Atomic States}.
\newblock \bibinfo{journal}{{\em Nature\/}} \bibinfo{volume}{612},
  \bibinfo{number}{7941} (\bibinfo{date}{Dec.} \bibinfo{year}{2022}),
  \bibinfo{pages}{661--665}.
\newblock
\showISSN{1476-4687}
\showDOI{%
\url{https://doi.org/10.1038/s41586-022-05363-z}}


\bibitem[\protect\citeauthoryear{Munro, Azuma, Tamaki, and Nemoto}{Munro
  et~al\mbox{.}}{2015}]%
        {munro_inside_2015}
\bibfield{author}{\bibinfo{person}{William~J. Munro}, \bibinfo{person}{Koji
  Azuma}, \bibinfo{person}{Kiyoshi Tamaki}, {and} \bibinfo{person}{Kae
  Nemoto}.} \bibinfo{year}{2015}\natexlab{}.
\newblock \showarticletitle{Inside Quantum Repeaters}.
\newblock \bibinfo{journal}{{\em IEEE J. Sel. Top. Quantum Electron.\/}}
  \bibinfo{volume}{21}, \bibinfo{number}{3} (\bibinfo{date}{May}
  \bibinfo{year}{2015}), \bibinfo{pages}{78--90}.
\newblock
\showISSN{1077-260X, 1558-4542}
\showDOI{%
\url{https://doi.org/10.1109/JSTQE.2015.2392076}}


\bibitem[\protect\citeauthoryear{Muralidharan, Li, Kim, L{\"u}tkenhaus, Lukin,
  and Jiang}{Muralidharan et~al\mbox{.}}{2016}]%
        {muralidharan_optimal_2016}
\bibfield{author}{\bibinfo{person}{Sreraman Muralidharan},
  \bibinfo{person}{Linshu Li}, \bibinfo{person}{Jungsang Kim},
  \bibinfo{person}{Norbert L{\"u}tkenhaus}, \bibinfo{person}{Mikhail~D. Lukin},
  {and} \bibinfo{person}{Liang Jiang}.} \bibinfo{year}{2016}\natexlab{}.
\newblock \showarticletitle{Optimal Architectures for Long Distance Quantum
  Communication}.
\newblock \bibinfo{journal}{{\em Sci. Rep.\/}} \bibinfo{volume}{6},
  \bibinfo{number}{1} (\bibinfo{date}{April} \bibinfo{year}{2016}),
  \bibinfo{pages}{20463}.
\newblock
\showISSN{2045-2322}
\showDOI{%
\url{https://doi.org/10.1038/srep20463}}


\bibitem[\protect\citeauthoryear{Nielsen and Chuang}{Nielsen and
  Chuang}{2010}]%
        {nielsen_quantum_2010}
\bibfield{author}{\bibinfo{person}{Michael~A. Nielsen} {and}
  \bibinfo{person}{Isaac~L. Chuang}.} \bibinfo{year}{2010}\natexlab{}.
\newblock \bibinfo{booktitle}{{\em Quantum Computation and Quantum
  Information\/} (\bibinfo{edition}{10th anniversary ed} ed.)}.
\newblock \bibinfo{publisher}{Cambridge University Press},
  \bibinfo{address}{Cambridge ; New York}.
\newblock
\showISBNx{978-1-107-00217-3}
\showLCCN{QA76.889 .N54 2010}


\bibitem[\protect\citeauthoryear{Pan, Bouwmeester, Weinfurter, and
  Zeilinger}{Pan et~al\mbox{.}}{1998}]%
        {pan_experimental_1998}
\bibfield{author}{\bibinfo{person}{Jian-Wei Pan}, \bibinfo{person}{Dik
  Bouwmeester}, \bibinfo{person}{Harald Weinfurter}, {and}
  \bibinfo{person}{Anton Zeilinger}.} \bibinfo{year}{1998}\natexlab{}.
\newblock \showarticletitle{Experimental Entanglement Swapping: Entangling
  Photons That Never Interacted}.
\newblock \bibinfo{journal}{{\em Phys. Rev. Lett.\/}} \bibinfo{volume}{80},
  \bibinfo{number}{18} (\bibinfo{date}{May} \bibinfo{year}{1998}),
  \bibinfo{pages}{3891--3894}.
\newblock
\showDOI{%
\url{https://doi.org/10.1103/PhysRevLett.80.3891}}


\bibitem[\protect\citeauthoryear{Pant, Krovi, Towsley, Tassiulas, Jiang, Basu,
  Englund, and Guha}{Pant et~al\mbox{.}}{2019}]%
        {pant_routing_2019}
\bibfield{author}{\bibinfo{person}{Mihir Pant}, \bibinfo{person}{Hari Krovi},
  \bibinfo{person}{Don Towsley}, \bibinfo{person}{Leandros Tassiulas},
  \bibinfo{person}{Liang Jiang}, \bibinfo{person}{Prithwish Basu},
  \bibinfo{person}{Dirk Englund}, {and} \bibinfo{person}{Saikat Guha}.}
  \bibinfo{year}{2019}\natexlab{}.
\newblock \showarticletitle{Routing Entanglement in the Quantum Internet}.
\newblock \bibinfo{journal}{{\em npj Quantum Inf.\/}} \bibinfo{volume}{5},
  \bibinfo{number}{1} (\bibinfo{date}{Dec.} \bibinfo{year}{2019}),
  \bibinfo{pages}{25}.
\newblock
\showISSN{2056-6387}
\showDOI{%
\url{https://doi.org/10.1038/s41534-019-0139-x}}


\bibitem[\protect\citeauthoryear{Pirandola, Bardhan, Gehring, Weedbrook, and
  Lloyd}{Pirandola et~al\mbox{.}}{2018}]%
        {pirandola_advances_2018}
\bibfield{author}{\bibinfo{person}{S. Pirandola}, \bibinfo{person}{B.~R.
  Bardhan}, \bibinfo{person}{T. Gehring}, \bibinfo{person}{C. Weedbrook}, {and}
  \bibinfo{person}{S. Lloyd}.} \bibinfo{year}{2018}\natexlab{}.
\newblock \showarticletitle{Advances in Photonic Quantum Sensing}.
\newblock \bibinfo{journal}{{\em Nat. Photon.\/}} \bibinfo{volume}{12},
  \bibinfo{number}{12} (\bibinfo{date}{Dec.} \bibinfo{year}{2018}),
  \bibinfo{pages}{724--733}.
\newblock
\showISSN{1749-4885, 1749-4893}
\showDOI{%
\url{https://doi.org/10.1038/s41566-018-0301-6}}


\bibitem[\protect\citeauthoryear{Pirandola, Eisert, Weedbrook, Furusawa, and
  Braunstein}{Pirandola et~al\mbox{.}}{2015}]%
        {pirandola_advances_2015}
\bibfield{author}{\bibinfo{person}{S. Pirandola}, \bibinfo{person}{J. Eisert},
  \bibinfo{person}{C. Weedbrook}, \bibinfo{person}{A. Furusawa}, {and}
  \bibinfo{person}{S.~L. Braunstein}.} \bibinfo{year}{2015}\natexlab{}.
\newblock \showarticletitle{Advances in Quantum Teleportation}.
\newblock \bibinfo{journal}{{\em Nat. Photon.\/}} \bibinfo{volume}{9},
  \bibinfo{number}{10} (\bibinfo{date}{Oct.} \bibinfo{year}{2015}),
  \bibinfo{pages}{641--652}.
\newblock
\showISSN{1749-4885, 1749-4893}
\showDOI{%
\url{https://doi.org/10.1038/nphoton.2015.154}}


\bibitem[\protect\citeauthoryear{Pompili, Delle~Donne, {te Raa}, {van der
  Vecht}, Skrzypczyk, Ferreira, {de Kluijver}, Stolk, Hermans, Pawe{\l}czak,
  Kozlowski, Hanson, and Wehner}{Pompili et~al\mbox{.}}{2022}]%
        {pompili_experimental_2022}
\bibfield{author}{\bibinfo{person}{M. Pompili}, \bibinfo{person}{C.
  Delle~Donne}, \bibinfo{person}{I. {te Raa}}, \bibinfo{person}{B. {van der
  Vecht}}, \bibinfo{person}{M. Skrzypczyk}, \bibinfo{person}{G. Ferreira},
  \bibinfo{person}{L. {de Kluijver}}, \bibinfo{person}{A.~J. Stolk},
  \bibinfo{person}{S.~L.~N. Hermans}, \bibinfo{person}{P. Pawe{\l}czak},
  \bibinfo{person}{W. Kozlowski}, \bibinfo{person}{R. Hanson}, {and}
  \bibinfo{person}{S. Wehner}.} \bibinfo{year}{2022}\natexlab{}.
\newblock \showarticletitle{Experimental Demonstration of Entanglement Delivery
  Using a Quantum Network Stack}.
\newblock \bibinfo{journal}{{\em npj Quantum Inf.\/}} \bibinfo{volume}{8},
  \bibinfo{number}{1} (\bibinfo{date}{Oct.} \bibinfo{year}{2022}),
  \bibinfo{pages}{121}.
\newblock
\showISSN{2056-6387}
\showDOI{%
\url{https://doi.org/10.1038/s41534-022-00631-2}}


\bibitem[\protect\citeauthoryear{Preskill}{Preskill}{2018}]%
        {preskill_quantum_2018}
\bibfield{author}{\bibinfo{person}{John Preskill}.}
  \bibinfo{year}{2018}\natexlab{}.
\newblock \showarticletitle{Quantum Computing in the NISQ Era and Beyond}.
\newblock \bibinfo{journal}{{\em Quantum\/}}  \bibinfo{volume}{2}
  (\bibinfo{date}{Aug.} \bibinfo{year}{2018}), \bibinfo{pages}{79}.
\newblock
\showISSN{2521-327X}
\showDOI{%
\url{https://doi.org/10.22331/q-2018-08-06-79}}


\bibitem[\protect\citeauthoryear{Proctor, Knott, and Dunningham}{Proctor
  et~al\mbox{.}}{2018}]%
        {proctor_multiparameter_2018}
\bibfield{author}{\bibinfo{person}{Timothy~J. Proctor},
  \bibinfo{person}{Paul~A. Knott}, {and} \bibinfo{person}{Jacob~A.
  Dunningham}.} \bibinfo{year}{2018}\natexlab{}.
\newblock \showarticletitle{Multiparameter Estimation in Networked Quantum
  Sensors}.
\newblock \bibinfo{journal}{{\em Phys. Rev. Lett.\/}} \bibinfo{volume}{120},
  \bibinfo{number}{8} (\bibinfo{date}{Feb.} \bibinfo{year}{2018}),
  \bibinfo{pages}{080501}.
\newblock
\showISSN{0031-9007, 1079-7114}
\showDOI{%
\url{https://doi.org/10.1103/PhysRevLett.120.080501}}


\bibitem[\protect\citeauthoryear{Ruf, Wan, Choi, Englund, and Hanson}{Ruf
  et~al\mbox{.}}{2021}]%
        {ruf_quantum_2021}
\bibfield{author}{\bibinfo{person}{Maximilian Ruf}, \bibinfo{person}{Noel~H.
  Wan}, \bibinfo{person}{Hyeongrak Choi}, \bibinfo{person}{Dirk Englund}, {and}
  \bibinfo{person}{Ronald Hanson}.} \bibinfo{year}{2021}\natexlab{}.
\newblock \showarticletitle{Quantum Networks Based on Color Centers in
  Diamond}.
\newblock \bibinfo{journal}{{\em J. Appl. Phys.\/}} \bibinfo{volume}{130},
  \bibinfo{number}{7} (\bibinfo{date}{Aug.} \bibinfo{year}{2021}),
  \bibinfo{pages}{070901}.
\newblock
\showISSN{0021-8979, 1089-7550}
\showDOI{%
\url{https://doi.org/10.1063/5.0056534}}


\bibitem[\protect\citeauthoryear{Schlosshauer}{Schlosshauer}{2019}]%
        {schlosshauer_quantum_2019}
\bibfield{author}{\bibinfo{person}{Maximilian Schlosshauer}.}
  \bibinfo{year}{2019}\natexlab{}.
\newblock \showarticletitle{Quantum Decoherence}.
\newblock \bibinfo{journal}{{\em Phys. Rep.\/}}  \bibinfo{volume}{831}
  (\bibinfo{date}{Oct.} \bibinfo{year}{2019}), \bibinfo{pages}{1--57}.
\newblock
\showISSN{03701573}
\showDOI{%
\url{https://doi.org/10.1016/j.physrep.2019.10.001}}


\bibitem[\protect\citeauthoryear{Shi and Qian}{Shi and Qian}{2020}]%
        {shi_concurrent_2020}
\bibfield{author}{\bibinfo{person}{Shouqian Shi} {and} \bibinfo{person}{Chen
  Qian}.} \bibinfo{year}{2020}\natexlab{}.
\newblock \showarticletitle{Concurrent Entanglement Routing for Quantum
  Networks: Model and Designs}. In \bibinfo{booktitle}{{\em Proceedings of the
  Annual Conference of the ACM Special Interest Group on Data Communication on
  the Applications, Technologies, Architectures, and Protocols for Computer
  Communication}}. \bibinfo{publisher}{ACM}, \bibinfo{address}{Virtual Event
  USA}, \bibinfo{pages}{62--75}.
\newblock
\showISBNx{978-1-4503-7955-7}
\showDOI{%
\url{https://doi.org/10.1145/3387514.3405853}}


\bibitem[\protect\citeauthoryear{Shor}{Shor}{1994}]%
        {shor_algorithms_1994}
\bibfield{author}{\bibinfo{person}{P.W. Shor}.}
  \bibinfo{year}{1994}\natexlab{}.
\newblock \showarticletitle{Algorithms for quantum computation: discrete
  logarithms and factoring}. In \bibinfo{booktitle}{{\em Proceedings 35th
  Annual Symposium on Foundations of Computer Science}}.
  \bibinfo{pages}{124--134}.
\newblock
\showDOI{%
\url{https://doi.org/10.1109/SFCS.1994.365700}}


\bibitem[\protect\citeauthoryear{Singh, Dev, Siljak, Joshi, and Magarini}{Singh
  et~al\mbox{.}}{2021}]%
        {singh_quantum_2021}
\bibfield{author}{\bibinfo{person}{Amoldeep Singh}, \bibinfo{person}{Kapal
  Dev}, \bibinfo{person}{Harun Siljak}, \bibinfo{person}{Hem~Dutt Joshi}, {and}
  \bibinfo{person}{Maurizio Magarini}.} \bibinfo{year}{2021}\natexlab{}.
\newblock \showarticletitle{Quantum Internet\textemdash Applications,
  Functionalities, Enabling Technologies, Challenges, and Research Directions}.
\newblock \bibinfo{journal}{{\em IEEE Commun. Surveys Tuts.\/}}
  \bibinfo{volume}{23}, \bibinfo{number}{4} (\bibinfo{year}{2021}),
  \bibinfo{pages}{2218--2247}.
\newblock
\showISSN{1553-877X, 2373-745X}
\showDOI{%
\url{https://doi.org/10.1109/COMST.2021.3109944}}


\bibitem[\protect\citeauthoryear{Steane}{Steane}{1998}]%
        {steane_quantum_1998}
\bibfield{author}{\bibinfo{person}{Andrew Steane}.}
  \bibinfo{year}{1998}\natexlab{}.
\newblock \showarticletitle{Quantum Computing}.
\newblock \bibinfo{journal}{{\em Rep. Prog. Phys.\/}} \bibinfo{volume}{61},
  \bibinfo{number}{2} (\bibinfo{date}{Feb.} \bibinfo{year}{1998}),
  \bibinfo{pages}{117--173}.
\newblock
\showDOI{%
\url{https://doi.org/10.1088/0034-4885/61/2/002}}


\bibitem[\protect\citeauthoryear{Van~Meter and Devitt}{Van~Meter and
  Devitt}{2016}]%
        {van_meter_path_2016}
\bibfield{author}{\bibinfo{person}{Rodney Van~Meter} {and}
  \bibinfo{person}{Simon~J. Devitt}.} \bibinfo{year}{2016}\natexlab{}.
\newblock \showarticletitle{The Path to Scalable Distributed Quantum
  Computing}.
\newblock \bibinfo{journal}{{\em Computer\/}} \bibinfo{volume}{49},
  \bibinfo{number}{9} (\bibinfo{date}{Sept.} \bibinfo{year}{2016}),
  \bibinfo{pages}{31--42}.
\newblock
\showISSN{0018-9162}
\showDOI{%
\url{https://doi.org/10.1109/MC.2016.291}}


\bibitem[\protect\citeauthoryear{Van~Meter, Satoh, Ladd, Munro, and
  Nemoto}{Van~Meter et~al\mbox{.}}{2013}]%
        {van_meter_path_2013}
\bibfield{author}{\bibinfo{person}{Rodney Van~Meter}, \bibinfo{person}{Takahiko
  Satoh}, \bibinfo{person}{Thaddeus~D. Ladd}, \bibinfo{person}{William~J.
  Munro}, {and} \bibinfo{person}{Kae Nemoto}.} \bibinfo{year}{2013}\natexlab{}.
\newblock \showarticletitle{Path Selection for Quantum Repeater Networks}.
\newblock \bibinfo{journal}{{\em Netw. Sci.\/}} \bibinfo{volume}{3},
  \bibinfo{number}{1} (\bibinfo{date}{Dec.} \bibinfo{year}{2013}),
  \bibinfo{pages}{82--95}.
\newblock
\showISSN{2076-0329}
\showDOI{%
\url{https://doi.org/10.1007/s13119-013-0026-2}}


\bibitem[\protect\citeauthoryear{Van~Meter, Touch, and Horsman}{Van~Meter
  et~al\mbox{.}}{2011}]%
        {van_meter_recursive_2011}
\bibfield{author}{\bibinfo{person}{Rodney Van~Meter}, \bibinfo{person}{Joe
  Touch}, {and} \bibinfo{person}{Clare Horsman}.}
  \bibinfo{year}{2011}\natexlab{}.
\newblock \showarticletitle{Recursive Quantum Repeater Networks}.
\newblock \bibinfo{journal}{{\em Prog. Informatics\/}} \bibinfo{number}{8}
  (\bibinfo{date}{March} \bibinfo{year}{2011}), \bibinfo{pages}{65}.
\newblock
\showISSN{1349-8614, 1349-8606}
\showDOI{%
\url{https://doi.org/10.2201/NiiPi.2011.8.8}}


\bibitem[\protect\citeauthoryear{Wehner, Elkouss, and Hanson}{Wehner
  et~al\mbox{.}}{2018}]%
        {wehner_quantum_2018}
\bibfield{author}{\bibinfo{person}{Stephanie Wehner}, \bibinfo{person}{David
  Elkouss}, {and} \bibinfo{person}{Ronald Hanson}.}
  \bibinfo{year}{2018}\natexlab{}.
\newblock \showarticletitle{Quantum Internet: A Vision for the Road Ahead}.
\newblock \bibinfo{journal}{{\em Science\/}} \bibinfo{volume}{362},
  \bibinfo{number}{6412} (\bibinfo{date}{Oct.} \bibinfo{year}{2018}),
  \bibinfo{pages}{eaam9288}.
\newblock
\showISSN{0036-8075, 1095-9203}
\showDOI{%
\url{https://doi.org/10.1126/science.aam9288}}


\bibitem[\protect\citeauthoryear{Wei, Jing, Zhang, Liao, Yuan, Fan, Lyu, Zhou,
  Wang, Deng, Song, Oblak, Guo, and Zhou}{Wei et~al\mbox{.}}{2022}]%
        {wei_towards_2022}
\bibfield{author}{\bibinfo{person}{Shi-Hai Wei}, \bibinfo{person}{Bo Jing},
  \bibinfo{person}{Xue-Ying Zhang}, \bibinfo{person}{Jin-Yu Liao},
  \bibinfo{person}{Chen-Zhi Yuan}, \bibinfo{person}{Bo-Yu Fan},
  \bibinfo{person}{Chen Lyu}, \bibinfo{person}{Dian-Li Zhou},
  \bibinfo{person}{You Wang}, \bibinfo{person}{Guang-Wei Deng},
  \bibinfo{person}{Hai-Zhi Song}, \bibinfo{person}{Daniel Oblak},
  \bibinfo{person}{Guang-Can Guo}, {and} \bibinfo{person}{Qiang Zhou}.}
  \bibinfo{year}{2022}\natexlab{}.
\newblock \showarticletitle{Towards Real-World Quantum Networks: A Review}.
\newblock \bibinfo{journal}{{\em Laser Photon. Rev.\/}} \bibinfo{volume}{16},
  \bibinfo{number}{3} (\bibinfo{year}{2022}), \bibinfo{pages}{2100219}.
\newblock
\showDOI{%
\url{https://doi.org/10.1002/lpor.202100219}}
\showeprint{https://onlinelibrary.wiley.com/doi/pdf/10.1002/lpor.202100219}


\bibitem[\protect\citeauthoryear{Wootters and Zurek}{Wootters and
  Zurek}{1982}]%
        {wootters_single_1982}
\bibfield{author}{\bibinfo{person}{W.~K. Wootters} {and} \bibinfo{person}{W.~H.
  Zurek}.} \bibinfo{year}{1982}\natexlab{}.
\newblock \showarticletitle{A Single Quantum Cannot Be Cloned}.
\newblock \bibinfo{journal}{{\em Nature\/}} \bibinfo{volume}{299},
  \bibinfo{number}{5886} (\bibinfo{date}{Oct.} \bibinfo{year}{1982}),
  \bibinfo{pages}{802--803}.
\newblock
\showISSN{1476-4687}
\showDOI{%
\url{https://doi.org/10.1038/299802a0}}


\bibitem[\protect\citeauthoryear{Yan, Zhou, Zhong, and Sheng}{Yan
  et~al\mbox{.}}{2023}]%
        {yan_advances_2023}
\bibfield{author}{\bibinfo{person}{Pei-Shun Yan}, \bibinfo{person}{Lan Zhou},
  \bibinfo{person}{Wei Zhong}, {and} \bibinfo{person}{Yu-Bo Sheng}.}
  \bibinfo{year}{2023}\natexlab{}.
\newblock \showarticletitle{Advances in Quantum Entanglement Purification}.
\newblock \bibinfo{journal}{{\em Science China Physics, Mechanics \&
  Astronomy\/}} \bibinfo{volume}{66}, \bibinfo{number}{5}
  (\bibinfo{date}{April} \bibinfo{year}{2023}), \bibinfo{pages}{250301}.
\newblock
\showISSN{1869-1927}
\showDOI{%
\url{https://doi.org/10.1007/s11433-022-2065-x}}


\bibitem[\protect\citeauthoryear{Yu, Lai, and Zhou}{Yu et~al\mbox{.}}{2021}]%
        {yu_protocols_2021}
\bibfield{author}{\bibinfo{person}{Nengkun Yu}, \bibinfo{person}{Ching-Yi Lai},
  {and} \bibinfo{person}{Li Zhou}.} \bibinfo{year}{2021}\natexlab{}.
\newblock \showarticletitle{Protocols for Packet Quantum Network
  Intercommunication}.
\newblock \bibinfo{journal}{{\em IEEE Trans. Quantum Eng.\/}}
  \bibinfo{volume}{2} (\bibinfo{year}{2021}), \bibinfo{pages}{1--9}.
\newblock
\showISSN{2689-1808}
\showDOI{%
\url{https://doi.org/10.1109/TQE.2021.3112594}}


\bibitem[\protect\citeauthoryear{Zhang and Zhuang}{Zhang and Zhuang}{2021}]%
        {zhang_distributed_2021}
\bibfield{author}{\bibinfo{person}{Zheshen Zhang} {and} \bibinfo{person}{Quntao
  Zhuang}.} \bibinfo{year}{2021}\natexlab{}.
\newblock \showarticletitle{Distributed Quantum Sensing}.
\newblock \bibinfo{journal}{{\em Quantum Sci. Technol.\/}} \bibinfo{volume}{6},
  \bibinfo{number}{4} (\bibinfo{date}{July} \bibinfo{year}{2021}),
  \bibinfo{pages}{043001}.
\newblock
\showDOI{%
\url{https://doi.org/10.1088/2058-9565/abd4c3}}


\bibitem[\protect\citeauthoryear{{\.Z}{\.Z}ukowski, Zeilinger, Horne, and
  Ekert}{{\.Z}{\.Z}ukowski et~al\mbox{.}}{1993}]%
        {zzukowski_event-ready-detectors_1993}
\bibfield{author}{\bibinfo{person}{M. {\.Z}{\.Z}ukowski}, \bibinfo{person}{A.
  Zeilinger}, \bibinfo{person}{M.~A. Horne}, {and} \bibinfo{person}{A.~K.
  Ekert}.} \bibinfo{year}{1993}\natexlab{}.
\newblock \showarticletitle{``Event-Ready-Detectors'' Bell Experiment via
  Entanglement Swapping}.
\newblock \bibinfo{journal}{{\em Phys. Rev. Lett.\/}} \bibinfo{volume}{71},
  \bibinfo{number}{26} (\bibinfo{date}{Dec.} \bibinfo{year}{1993}),
  \bibinfo{pages}{4287--4290}.
\newblock
\showDOI{%
\url{https://doi.org/10.1103/PhysRevLett.71.4287}}


\end{thebibliography}

\end{document}